\documentclass{pasa}%

\usepackage{graphicx}
\usepackage{tablefootnote}
\usepackage{svg}
\usepackage{listings}
\usepackage{appendix}
\usepackage{blindtext}
\usepackage[normalem]{ulem}
\usepackage{cancel}
\usepackage{footnote}
\usepackage{dblfloatfix}
\usepackage{caption}

\newcommand{\expa}{Experimental Astronomy}
\newcommand{\icarus}{Icarus}

\colorlet{VIOLET}{violet}
\newcommand\sw[1]{{\fontfamily{lmss}\selectfont\color{violet}#1}}

\newcommand\email[1]{{\color{blue}#1}}

\title[\sw{SDtracker}: Spacecraft Doppler tracker]{High spectral resolution multi-tone Spacecraft Doppler tracking software:\\ Algorithms and implementations}

\author[Molera Calvés et al.]{G. Molera Calvés$^1$\thanks{\email{guifre.moleracalves@utas.edu.au}}~, S.~V. Pogrebenko$^2$, J.~F. Wagner$^3$, G. Cim\`o$^2$, L.~I. Gurvits$^{2,4}$, T.~M. Bocanegra-Baham\'on$^5$, D.~A. Duev$^6$, N.~V. Nunes$^7$

\affil{$^1$Physics discipline, School of Natural Sciences, University of Tasmania, Private Bag 37, Hobart, TAS 7000, Australia}
\affil{$^2$Joint Institute for VLBI ERIC, Dwingeloo, Oude Hogeveensedijk 4, Dwingeloo 7991PD, The Netherlands}
\affil{$^3$Max Planck Institute for Extraterrestrial Physics, Giessenbachstr. 1, Garching, 85748, Germany} 
\affil{$^4$ Aerospace Faculty, Delft University of Technology, Kluyverweg 1, Delft 2629HS, The Netherlands}
\affil{$^5$Jet Propulsion Laboratory, California Institute of Technology, 4800 Oak Grove Dr., Pasadena, CA 91125, USA} 
\affil{$^6$Division of Physics, Mathematics, and Astronomy, California Institute of Technology, Pasadena, CA 91125, USA} 
\affil{$^7$York University, 4700 Keele St, Toronto, ON M3J 1P3, Canada}
}

\jid{PASA}
\jyear{\the\year}

\usepackage{aas_macros}
\usepackage{hyperref} 
\hypersetup{colorlinks,citecolor=blue,linkcolor=blue,urlcolor=blue}

\begin{document}

\begin{frontmatter}
\maketitle

\begin{abstract}
We present a software package for single-dish data processing of spacecraft signals observed with VLBI-equipped radio telescopes. The Spacecraft Doppler tracking (\sw{SDtracker}) software allows one to obtain topocentric frequency detections with a sub-Hz precision, and reconstructed and residual phases of the carrier signal of any spacecraft or landing vehicle at any location in the Solar System. These data products are estimated using the ground-based telescope's highly stable oscillator as a reference, without requiring an a priori model of the spacecraft dynamics nor the downlink transmission carrier frequency. The software has been extensively validated in multiple observing campaigns of various deep space missions and is compatible with the raw sample data acquired by any standard VLBI radio telescope worldwide. In this paper, we report the numerical methodology of \sw{SDtracker}, the technical operations for deployment and usage, and a summary of use cases and scientific results produced since its initial release.
\end{abstract}

\begin{keywords}
spacecraft tracking -- Very Long Baseline Interferometry -- radio telescopes -- Doppler -- software correlator -- radio science -- single-dish -- space debri
\end{keywords}
\end{frontmatter}

\section{INTRODUCTION}
\label{sec:intro}

Radio techniques are an indispensable part of the instrumentation exploited by space missions for trajectory and orbit determination, navigation, commanding, and data exchange. Traditionally, these components of space mission operations and various science applications are provided by dedicated deep space tracking facilities, such as the NASA's Deep Space Network (DSN) and the European Space tracking network (Estrack). Radio astronomy facilities, in particular those involved in Very Long Baseline Interferometry (VLBI), offer a powerful complement to nominal deep space tracking assets. Modern VLBI networks involve more than $50$ radio telescopes distributed globally. These telescopes can provide additional support to radio science experiments and assist in the measurement of spacecraft state vectors exploiting dedicated software tools.

The methodological approach presented in this paper descends from the VLBI tracking experiment conducted with the Venus atmosphere balloons of the VEGA mission in $1985$. The mission employed single-dish Doppler tracking with VLBI antennas, as in this paper, and included interferometric analysis based on differential fringe rates~\citet{Sagdeev1986,Sagdeev1992}. The software algorithms were further developed in support to the Doppler Wind Experiment of the ESA's Huygens Probe mission~\citet{Lebreton2005,Bird2005}. In the latter experiment, multiple VLBI radio telescopes around the world simultaneously tracked the signal of the probe during its parachute descent and soft landing on the surface of Titan on $2005.01.14$. Data recorded by all the antennas were transferred to the Joint Institute for VLBI in Europe (JIVE) for processing. The Huygens VLBI tracking experiment provided important support to the mission. It demonstrated the potential of radio astronomy techniques being used for tracking of deep space missions, namely, the use of radio interferometry for plane of sky motions, and radio Doppler measurements for Earth-relative velocities~\citet{Witasse2006}. These achievements triggered the development of a versatile software package to support spacecraft tracking using VLBI radio telescopes and streamline the data processing of the spacecraft carrier signal for radio science purposes. The first version of the software was completed in $2010$ to explore the capabilities of VLBI radio telescopes and the computational performance of the available data processing equipment.

The Spacecraft Doppler tracking (\sw{SDtracker}) software was developed in order to augment the established nominal means of getting accurate astrometry and precise Doppler measurements of deep space missions in the Solar System in almost real-time. The astrometric component of this task, not itself part of \sw{SDtracker}, is based on the use of the VLBI technique and provides the most accurate estimates (at the level of sub-milliarceconds or nanoradians) of celestial coordinates of natural radio sources. However, the instrumentation of individual radio telescopes involved in VLBI observations permit more than the usual interferometric processing of received signals, in particular, it allows spectrometric measurements as well. For the Solar System objects, this has been demonstrated, e.g., by detection of water maser emission in the Saturnian system~\citet{Pogrebenko2009}. One of the most important applications of spectrometric analysis of spacecraft radio signals, implemented in \sw{SDtracker}, is measurement of the signal's Doppler shift thus enabling estimates of the spacecraft's radial velocity. The VLBI data (lateral coordinates) and Doppler shift (radial velocity) provide complementary inputs into the determination of spacecraft state vectors thus supporting navigation techniques.

JIVE initiated development of Planetary Radio Interferometry and Doppler Experiment (PRIDE) in $2010$ aiming to utilise VLBI instrumentation for tracking planetary spacecraft in the Solar System. PRIDE uses the phase referencing VLBI technique together with topocentric Doppler measurements from multiple antennas with high precision in the interests of diverse scientific outcomes. These included the highly accurate orbit determination (OD) of the ESA's Venus Express (VEX) spacecraft~\citet{Duev2012}; the characterisation of the interplanetary plasma and the total electron content along the VEX line of sight~\citet{Molera2014}; the improvement of the OD accuracy of the space VLBI mission RadioAstron~\citet{Duev2015}; the tracking of a close flyby of Phobos by the ESA's Mars Express (MEX)~\citet{Duev2016} and the formulation of a generalised Doppler tracking and noise budget in that test case~\citet{Bocanegra2018}; the detection of a coronal mass ejection along the MEX line of sight~\citet{Molera2017}; and the study of the Venusian atmosphere by means of VEX radio occultations~\citet{Bocanegra2019}. Furthermore, a forward look for PRIDE participation in the ESA's Jupiter Icy Moons Explorer (JUICE) mission are discussed by~\citet{Dirkx2017}. A complementary support by PRIDE to the lander radio science (LaRa) experiment of the ESA-Roscosmos ExoMars–$2022$ mission has been considered too~\citet{Dehant2020}. PRIDE methodology proved to be useful for fundamental physics experiments on verification of the Einstein Equivalence Principle~\citet{Litvinov2018,Nunes2020,Litvinov2021}. All these applications rely on the \sw{SDtracker} software described in this paper.

This paper presents the principles of the \sw{SDtracker} single-dish spectral processing software for VLBI antenna data, and highlights its use in specific scientific applications. In Section~\ref{algorithms}, we introduce the software components, as well as their algorithmic background. We also include the description of the input data, instructions for the users and the resulting data products. In Section~\ref{featperf}, we highlight the relevant features and the best practice for conducting observations of spacecraft. This section also includes an assessment of the computational performance and benchmarks. In Section~\ref{validation}, we show examples of the data analysis at different stages of the processing pipeline and demonstrate how the software is applied in various scientific scenarios. Sections~\ref{outlook}~\&~\ref{conclusions} conclude the paper with a discussion of the relevant scientific areas that this software will aim to, as well as charters the next steps in its development.

\section{Software components and algorithms}
\label{algorithms}

\begin{figure}[hbtp]
 \centering
 \includegraphics[width=1.0\columnwidth]{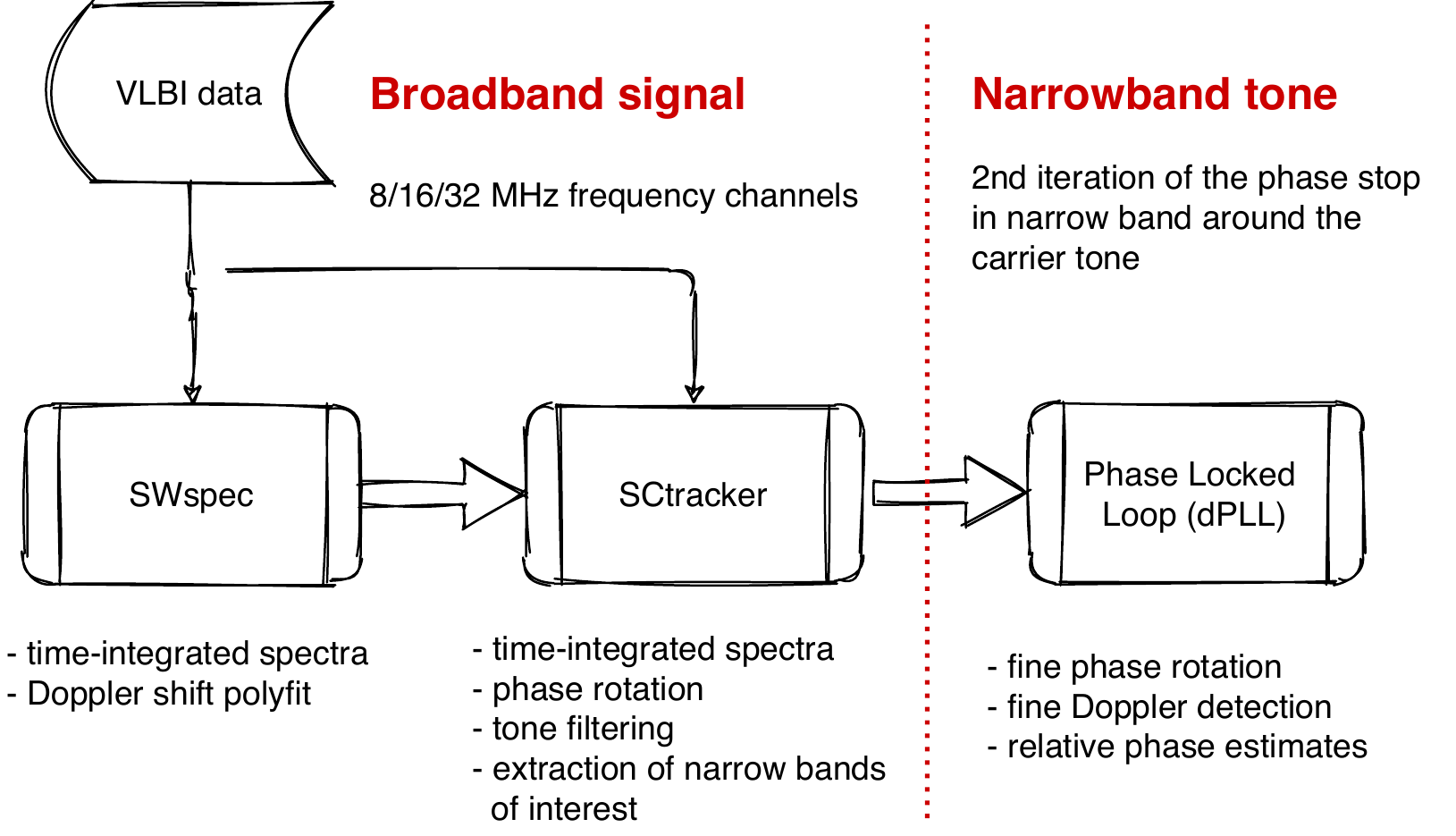}
 \caption{The spacecraft signal observed by a VLBI–equipped radio telescope is processed by three software components: \sw{SWspec}, \sw{SCtracker} and \sw{dPLL}. The main products of \sw{SDtracker} are the topocentric frequency detections and the residual phase of the spacecraft carrier signal.}
 \label{fig:data_processing}
\end{figure}

\sw{SDtracker} consists of three distinct open source packages that process the data acquired by a VLBI-equipped radio telescope to generate several data products for radio science and space missions support. The VLBI data format is, however, different from what is used in deep space tracking and communication systems, such as DSN and Estrack. While the latter provide frequency measurements with respect to the initial transmitted tone~\citet{ccsds2019}, radio astronomical systems measure the topocentric frequency and residual phase of the spacecraft carrier radio signal recorded in a broader bandwidth (several hundred megahertz to gigahertz). The full data processing pipeline, as illustrated in Figure~\ref{fig:data_processing}, consists of:

\begin{itemize}
 \item Software spectrometer (\href{https://gitlab.com/gofrito/swspec/}{SWspec})\footnote{~\url{gitlab.com/gofrito/swspec/}; accessed on $2021.04.06$.}
 \item Multi-tone spacecraft tracker (\href{https://gitlab.com/gofrito/sctracker/}{SCtracker})\footnote{~\url{gitlab.com/gofrito/sctracker/}; accessed on $2021.04.06$.}
 \item Digital Phase Locked Loop (\href{https://gitlab.com/gofrito/pysctrack/}{dPLL})\footnote{~\url{gitlab.com/gofrito/pysctrack/}; accessed on $2021.04.06$.}
\end{itemize}

\sw{SWspec} computes a time-series of the signal power spectra in the whole available band (usually several megahertz), with a selectable spectral resolution  from several kHz to sub-Hz. \sw{SWspec} provides initial detection and estimation of the Doppler shift and its variation with time. It can also be used as a classical astronomical spectrometer for the study of natural radio sources such as astronomical masers (see Section~\ref{singledish}.

\sw{SCtracker} uses the time-series of initial frequency detections from \sw{SWspec} to stop the varying Doppler shift to first degree of accuracy and cut out a narrow frequency band (or several bands) around the frequency('s) of interest, thus providing input for the next step of signal processing. 

\sw{dPLL} conducts a second iteration of phase-stopping based on the output data from \sw{SCtracker}. It performs more precise frequency detections in a narrow band by compensating for residual phase rotation. \sw{dPLL} outputs time-series of the signal amplitude, frequency and phase, and the signal itself in a further narrowed down frequency band. A detailed description of the algorithms used in these routines is presented below.

\subsection{Software spectrometer (\sw{SWspec})}
\label{swspec}

\sw{SWspec} calculates the spectral power integrated over a selected time of broadband radio signals. It can be utilised to estimate the fast Doppler variations of spacecraft carrier signals as well as single-dish spectral line observations.

\subsubsection{Astronomical software spectrometer}

\sw{SWspec} performs the initial detection of the spacecraft carrier signal, subcarriers and ranging tones (if present) at any desired spectral resolution. It is based on the software spectrometer developed for processing observations of the ESA's Huygens Titan Probe and the SMART-1 Lunar mission~\citet{Pogrebenko2006}. \sw{SWspec} is a versatile and highly-configurable spectrometer, written in \verb!C++! with a \verb!Python! wrapper, that handles radio astronomical data recorded by VLBI data acquisition systems (DAS). 

\sw{SWspec} is compatible with most current and legacy VLBI data formats. It utilises a common external library (\sw{mark5access})\footnote{\url{svn.atnf.csiro.au/difx/libraries/mark5access/};\\ accessed on 2021.04.28}, to read the quantised time domain signal recorded by the VLBI DAS and extracts any selected frequency channel. It then performs an accurate window-overlap-add (WOLA) discrete Fourier transform (DFT) per iteration and calculates the time-integrated spectrum. The user can set a wide range of spectral resolution (from sub-Hz to tens of kHz) and integration times (from seconds to minutes).

Both \sw{SWspec} and \sw{SCtracker} use window (apodisation) functions to improve the reliability of detections, smooth the spectrum, and reduce the aliasing and cyclicity effects. The choice of apodisation function is a compromise between spectral leakage effects and spectral resolution. The software use a large number of spectral points ($\ge10^6$ points) to compensate for the spectral smearing. The available window functions ($w[n]$) in the software and their corresponding main lobe full width at half maximum (FWHM) are listed in Table~\ref{tab:windows}. Note that the cosine squared and Hann windows are mathematically the same, but they are implemented differently in the software.

\begin{table}[!hbt]
  \caption{Available window functions ($w[n]$) in \sw{SWspec} and \sw{SCtracker}, their numerical basis, and main lobe full width at half maximum (FWHM) in spectral bins. Where $N$ is the length of the data vector in samples and $n$ is the sample number (starting with $0$).}
  \label{tab:windows}
  \small
  \tabcolsep=0.08cm
  \begin{tabular}{ l | l c }
  \hline\hline
    Window & Description ($w[n]$) & FWHM \\
    \hline
    Cosine$^{*}$ & $\sin \left( \frac{\pi n}{N-1}\right)$ & 1.64 \\
    Cos square & $1 - \cos^{2}\left( \frac{\pi n}{N-1}\right)$& 2.00 \\
    Hann       & $0.5 - 0.5 \cos\left( \frac{2 \pi n}{N-1}\right)$ & 2.00 \\
    Hamming    & $0.54 - 0.46 \cos\left(\frac{2 \pi n}{N-1}\right)$ & 1.82 \\
    Blackman   & $0.42 - 0.5 \cos\left( \frac{2 \pi n}{N - 1}\right) +  0.08 \cos\left( \frac{4 \pi n}{N -1}\right)$ & 2.30 \\
    \hline\hline
  \end{tabular}
  \tabnote{$^{*}\,$The cosine function becomes a sinus because of our definition of $n$, which starts with $0$.}
\end{table}

Each of the window functions provides advantages and disadvantages. For WOLA applications, it is desirable that the window is (a)~as close as possible to amplitude $0$ at the start and end of an interval, and close to $1$ at its middle; and (b)~the sum of the windows shifted by half the length of the data vector be close to flat $1$. For example, the cosine window does not obey the rule (b)~causing modulation of the output power. We selected the cosine squared (or the Hann) window function to be used by default to calculating the DFT of spacecraft signals after extensive testing.

\sw{SWspec} is configured by a control file that initialises all the input parameters described in Table~\ref{tab:ini_swspec}. A number of these parameters depend on the antenna setup used during the observations: the bandwidth of each critically sampled frequency channel (BandwidthHz), the data format (SourceFormat), and the number of frequency channels (SourceChannels). Other parameters, such as the number of spectral points (FFTpoints), the integration time (FFTIntegrationTime) and the FFT overlap (FFToverlapFactor) are used to adjust the spectral resolution and the accuracy of the carrier detection. These depend mainly on the radial acceleration of the spacecraft. The frequency resolution (dF) is equivalent to the sampling rate (SR) divided by the number of spectral points (dF$\,=\,$SR$\,\cdot\,$FFTpoints$^{-1}$) in Hz, where SR is twice the bandwidth of the frequency channel (SR$\,=\,2\,\cdot\,$BW). The time resolution (dT) is equal to the integration time of each spectrum in seconds.

\begin{table*}[!hbt]
 \caption{List of all the parameters available in the control file to initialise and run the software spectrometer.}
 \label{tab:ini_swspec}
 \begin{tabular}{l | l c }
 \hline\hline
 \textbf{Parameter} & \textbf{Definition} & \textbf{Default values}\\
 \noalign{\smallskip}\hline\noalign{\smallskip}
 FFTpoints             & number of bins in the spectral output & 3.2e6\\
 SourceFormat          & input data format & VDIF \\
 SourceChannels        & number of frequency channels in the raw data & 4\\
 BandwidthHz           & bandwidth of the frequency channel(s) & 8e6 \\
 UseChannel1           & select first channel source & 1\\
 UseChannel2           & select second channel source & 2\\
 DoCrossPol            & calculate cross-polarisation product between channels 1 and 2 & No \\
 ExtractPcal           & extract Phase Cal tone (Boolean) & No \\
 FFTIntegrationTime    & integration time in seconds for FFT average & 5\\ 
 FFToverlapFactor      & number of FFT's to overlap & 2\\
 \hline
 BaseFilename1         & binary spectral output filename 1 \\
 BaseFilename2         & binary spectral output filename 2 \\
 \noalign{\smallskip}
 \hline\hline
 \end{tabular}
\end{table*}

The number of spectral points and the integration time are the most critical parameters for reliable detection of the spacecraft signal. The selection of the appropriate parameters depends on the receiving antenna system equivalent flux density (SEFD), the spacecraft transmission power, propagation losses, and the relative motion between the ground-station and spacecraft. These parameters can differ significantly if the target is an orbiter, a lander or an interplanetary spacecraft. More details on the selection of these parameters is provided in Section~\ref{analysis}. The spectral output data are written to disk in $32$-bit binary format. These spectra are used for modelling the Doppler shift between the ground-station and spacecraft.

\subsubsection{Phase polynomial fit}

A series of \verb|Python|~\citep{Van1995} libraries have been developed to visualise the spectral data and extract intrinsic measurements of the radio signal(s). These scripts are available in the \sw{pysctrack} Git repository. \sw{CalculateCpp.py} allows the user to detect the frequency of the main carrier signal, estimate the rate of the Doppler shift and build the phase polynomial fit.

\sw{CalculateCpp.py} searches for the peak of the spacecraft carrier signal and estimates the topocentric frequency detections along the entire frequency band and scan length. When \sw{SWspec} is used with high spectral resolution the spectrum of the carrier line appears shifted by the Doppler shift. Typical processing configuration consists of $3.2\cdot10^{6}$ FFTpoints for a $16$\,MHz frequency channel. To compensate for the spectral power being spread out across multiple bins (smearing), the actual location of the carrier tone is estimated via a spectral centroid over five frequency bins around the peak. Thus, the spectral line position is achieved with sub-bin precision.

The time-series of topocentric frequency detections $f(t)$ of the spacecraft carrier signal are used to construct a frequency polynomial regression fit $F(t)$ of $m$-th using the weighted least mean square method. 

\begin{equation}
 F(t) = C_{\rm fs}(0) + C_{\rm fs}(1) \cdot t + ... + C_{\rm fs}(m-1) \cdot t^{m-1},
 \label{eq:polyfit}
\end{equation}

\noindent where $C_{\rm fs}$ are the frequency polynomial coefficients. The degree of polynomial ($m$) ranges between $3$ and $7$ depending on the length of the scan, the spectral resolution and the rate change of the Doppler shift. We use the signal-to-noise ratio (SNR) as weights per sample.

SNR for each sample can be determined as a ratio of detected spectral line power to the rms noise power in the near-by region of spectrum, defined as:

\begin{equation}
    {\rm SNR} = \frac{P_{\rm sc}}{k \cdot T_{\rm sys} \cdot {\rm dF}},
    \label{eq:SNR}
\end{equation}

\noindent where $k$ is the Boltzmann constant, $T_{\rm sys}$ the system temperature of the antenna measured in [K], $P_{\rm sc}$ the spacecraft tone spectral power in [W], and dF the frequency resolution in [Hz] used in \sw{SWspec}. In the case when it is not possible to compute the rms noise level, weights can be set to 1. See the \sw{pysctrack} user guide in the GitLab repository\footnote{~\url{gitlab.com/gofrito/pysctrack/-/wikis/}; accessed on 2021.04.15} for more information.


The phase polynomial
\begin{equation}
 P(t) = C_{\rm ps}(0) + C_{\rm ps}(1) \cdot t + ... + C_{\rm ps}(m-1) \cdot t^{m-1},
 \label{eq:phasefit}
\end{equation}

\noindent is constructed by the phase polynomial coefficients ($C_{\rm ps}$) calculated from the frequency polynomial by integrating its coefficients (for m$\,>\,0$) according to:
\begin{equation}
 C_{\rm ps}(m) = \frac{2 \pi}{m} \cdot C_{\rm fs}(m-1),
 \label{eq:cps}
\end{equation}

\noindent where the initial phase (integration constant) is set to $0$, the second is the phase rate [rad/s], the third is the acceleration [rad/s$^2$], and so on; and $C_{\rm fs}$ are the frequency polynomial coefficients per second being the first component the detected tone frequency [Hz], rate [Hz/s], acceleration [Hz/s$^2$]. For computational convenience the phase coefficients are transformed from phase per second to phase per sample ($C_{\rm pp}$) as:
\begin{equation}
 C_{\rm pp}(m) = C_{\rm ps}(m) \cdot SR^{\,-m}.
 \label{eq:cpp}
\end{equation}

\noindent where $SR$ is the sampling rate in samples per second.


The procedure \sw{CalculateCpp.py} performs the tasks described above. It outputs the SNR, the topocentric frequency detections, the stochastic Doppler noise and stores them in the \textit{Fdets} file. A detailed description of the output files is presented in Section~\ref{data_products}. The script also generates the $C_{\rm pp}$ file that contains the phase polynomial coefficients with respect to the initial base band frequency channel in terms of radians per sample and the $C_{\rm fs}$ file with the frequency polynomial coefficients in terms of Hz per second.

The $C_{\rm pp}$ and $C_{\rm fs}$ coefficients per are used as input parameters for \sw{SCtracker}.

\subsection{Multi-tone spacecraft tracker (\sw{SCtracker})}
\label{sctrack}

\sw{SCtracker} is the core package for tracking the carrier radio signals transmitted by spacecraft. This program is an evolution of the software developed for tracking of the Huygens Probe~\citet{Pogrebenko2004}, and is the prototype of the current European VLBI Network (EVN) Super FX correlator (SFXC)~\citet{Keimpema2015}. \sw{SCtracker} is written in \verb!C++! with a light graphical user interface in \verb!Python!. Table~\ref{tab:3} lists the software's main input and output files. All the configurable parameters in the \sw{SCtracker} control file are described in the Table~\ref{tab:appendixSCtrackerConfig} at the Appendix.

\begin{table*}[hbt]
\center
 \caption{Main input files and settings required by \sw{SCtracker}, and the output products after execution. All files are stored in the same directory in which \sw{SCtracker} runs.}
 \label{tab:3}
 \begin{tabular}{l | l}
 \hline\hline\noalign{\smallskip}
 \textbf{File} & \textbf{Description} \\
 \noalign{\smallskip}\hline\noalign{\smallskip}
 \textbf{input files} : & \\
 raw file   & raw data recorded by a VLBI radio telescope \\
 tones file & ranging tones separated by frequency offset from the carrier\\
 ini file   & initial settings file, similar to \sw{SWspec} configuration file \\
 $C_{\rm pp}$   & phase polynomial coefficients \\
 $C_{\rm fs}$   & frequency polynomial coefficients \\
 \noalign{\smallskip}\hline
 \textbf{output files} : &  \\
 *\_phasevalues.txt & relative phase values of the tone - low resolution\\
 *\_runlog.txt      & debug information from \sw{SCtracker}\\
 *\_scspec.bin      & spectral data after the phase-stop\\
 *\_starttiming.txt & initial time as read from VLBI data headers\\
 *\_tone0.bin       & narrow band filtered spacecraft tone\\
 *\_toneX.bin       & X-number narrow band filtered spacecraft tone\\ 
 *\_tonebinning.txt & initial bin of the spacecraft tone\\
 \hline\hline
 \end{tabular}
\end{table*}

\sw{SCtracker} utilises the estimated phase polynomial coefficients ($C_{\rm ps}$) to conduct an initial phase-stop by doing a phase-rotation of the entire recorded bandwidth. The desired carrier frequency tone is then selected to perform signal filtering, extract a narrow band around the tone with the selected bandwidth, and detect the relative phase of the tone. \sw{SCtracker} is capable of simultaneously tracking the spacecraft carrier signal, its subcarriers and all the ranging tones. This feature is described in more detail in Section~\ref{dor}.

\sw{SCtracker} transforms the initial baseband signal $S(t)$ to a complex signal $S'(t)$ with the nonlinear part of the phase evolution corrected as expressed in:

\begin{equation}
 S'(t) = S(t) \cdot e^{-i \phi_{\rm corr(t)}}.
\label{eq:analytical}
\end{equation}

Equation~\ref{eq:analytical} is the analytical form of the double-precision polynomial transformation which is applied to the baseband sample sequence ($x[n]$), where the applied phase correction ($\phi_{\rm corr(t)}$) corresponds to the polynomial fit coefficients ($C_{\rm pp}$) calculated previously.

\begin{equation}
  x'[n]=x[n]\exp \left( \pm i \sum\limits_{k=1}^{m-1}C_{\rm pp}(k)\cdot T_{n}^{\,k}\right),
  \label{SCtracker}
\end{equation}

\noindent where $x'[n]$ are the new samples, $x[n]$ are the original raw samples, $T_{n}$ is the time relative to the beginning of the data segment, and $\pm$ signs take into account that the frequency channel can be upper or lower side band. The calculated time-integrated window-overlapped spectra of the phase stopped baseband signal are saved to disk. These use the same spectral representation as produced with \sw{SWspec}, with the frequency shift corrected as specified by the polynomial approximation according to Equation~\ref{SCtracker}. The spectral output of \sw{SCtracker} shows the spacecraft carrier signal as a single spectral line along the entire spectral time series.

A narrow band around each tone is stored for further signal processing at a higher resolution. The user can set the bandwidth of the output tone in the initial control file. Typically, a band of $2$\,kHz is used. The narrow bands are extracted directly from the stopped baseband signal around each of the frequency tones specified in the tones file. Each tone is associated with a number, $0$ being the carrier line, and $1$ onwards the subsequent ranging tone(s). They are sorted based on their proximity to the spacecraft carrier line. 

The current implementation of the software allows an unlimited number of extracted narrow bands, with an arbitrary distribution in the input band. These bands (with the tones in the centre) are filtered out into continuous complex time-domain signals using a $2^{\rm nd}\,$order WOLA DFT-based algorithm of the Hilbert transform approximation. The extracted signals in temporal domain are written to output files in binary format for further post-processing. All complex data are stored with floating-point precision to avoid losing resolution between the different iterations. The initial timestamp of the processed scan, based on the VLBI data header, and the cut-off frequencies of the narrow band(s) are written to auxiliary output files.

\subsection{Digital Phase Lock Loop (\sw{dPLL})}
\label{pll}

\sw{dPLL} provides a finer detection of the topocentric frequency measurements, an ultra-narrow band of the carrier signal, and the residual phase with respect to the station clock. It repeats with high precision the steps described in Sections~\ref{swspec} and~\ref{sctrack} on the filtered narrow band spacecraft signal. The \sw{dPLL.py} script that performs these operations, can be found in the \sw{pysctrack} software package.

The first step in \sw{dPLL} consists of calculating the new time-integrated overlapped spectra with longer integration time and finer spectral resolution. It estimates a regression line to the frequency detections and calculates a new set of phase polynomial coefficients, $C_{\rm pp}$(\verb!r2i!), using the WLMS method. A band-pass filter centred at the tone is applied, followed by a Hilbert transform approximation to discard a negative part of the spectrum. The user can select the output filter of \sw{dPLL} via command line. By default, the output bandwidth (BW$_{\rm o}$) is set to $20\,$Hz.

The second step in \sw{dPLL} consists of shifting the spacecraft tone to a lower frequency, near the baseband cut-off. We set F$_{\rm startc}$ as the first element of C$_{\rm fs}$. The bin position in the initial narrow band is defined as F$_{\rm max}$. We set the frequency target (F$_{\rm target}$) in the middle of the output bandwidth, thus, the carrier signal can be phase rotated with the frequency shift of F$_{\rm rot}$ = (F$_{\rm max}$ - F$_{\rm target}$). In order to perform the phase-stop in an ultra narrow band, we apply the same concept as in Equation~\ref{eq:analytical}:


\begin{equation}
  s_{\rm sc}(t) = s(t) \cdot e^{-i 2 \pi F_{\rm rot} \cdot t}.
\end{equation}

The phase of the spacecraft signal is retrieved from the signal in the time domain after accounting for the phase unwrap. The phase follows an almost straight trend with a slope proportional to $2 \pi F_{\rm target}$. Finally, the residual phase of the signal with respect to the station clock is extracted by applying a new phase polynomial fit using again a WLMS method. The order of the third polynomial fit is mostly consistent with the two previous fits.

The reconstructed phase of carrier signal is one of the outputs of \sw{dPLL.py}. The phase is a function of the residual phase $\delta\Phi(t)$, the three phase-rotations performed inside \sw{dPLL} ($F_{\rm pll}$ = $F_{\rm target}$ + $F_{\rm rot}$ + $F_{\rm startc}$) and the \sw{SCtracker} phase-stop:

\begin{equation}
\label{eq:phase}
\begin{split}
  \Phi(t) = - \delta\Phi(t) + 2 \pi F_{\rm pll} \cdot t + \Phi_{\rm sctrack}(t),
\end{split}
\end{equation}

\noindent where $\Phi_{sctrack}(t)$ is the polynomial model obtained from the raw Doppler measurements after running \sw{SWspec} and \sw{SCtracker}, and can be derived as:

\begin{equation}
 \Phi_{\rm sctrack}(t) = \sum_{\kappa=1}^{N_{\rm pp}} \left[C_{\rm pp,ip} \cdot \left(\frac{t}{T_{\rm span}} \right)^{\kappa} \right],
\end{equation}

\noindent where $N_{\rm pp}$ is the highest polynomial order used in the \sw{SCtracker} and \sw{dPLL}, $C_{\rm pp}$ is the sum of the respective phase polynomial coefficients ($C_{\rm pp1}$ and $C_{\rm pp2}$), and $T_{\rm span}$ is the length of the full scan.

This second step can be repeated multiple times in order to achieve a higher precision in the tone detection by narrowing the final band-pass filter. For studies of interplanetary scintillation, as described in~\citet{Molera2014}, data were filtered down to a bandwidth of $20\,$Hz. However, when radio signals propagated through the solar corona and were affected by strong scintillation, data were analysed in a larger bandwidth (i.e. $50\,$Hz). For phase referencing and orbit determination, the resolution of the carrier signal is improved by filtering down to a bandwidth of $5$ or $1\,$Hz.

\subsection{Data products}
\label{data_products}

\begin{figure}[hbt]
    \centering
    \includegraphics[width=1.0\columnwidth]{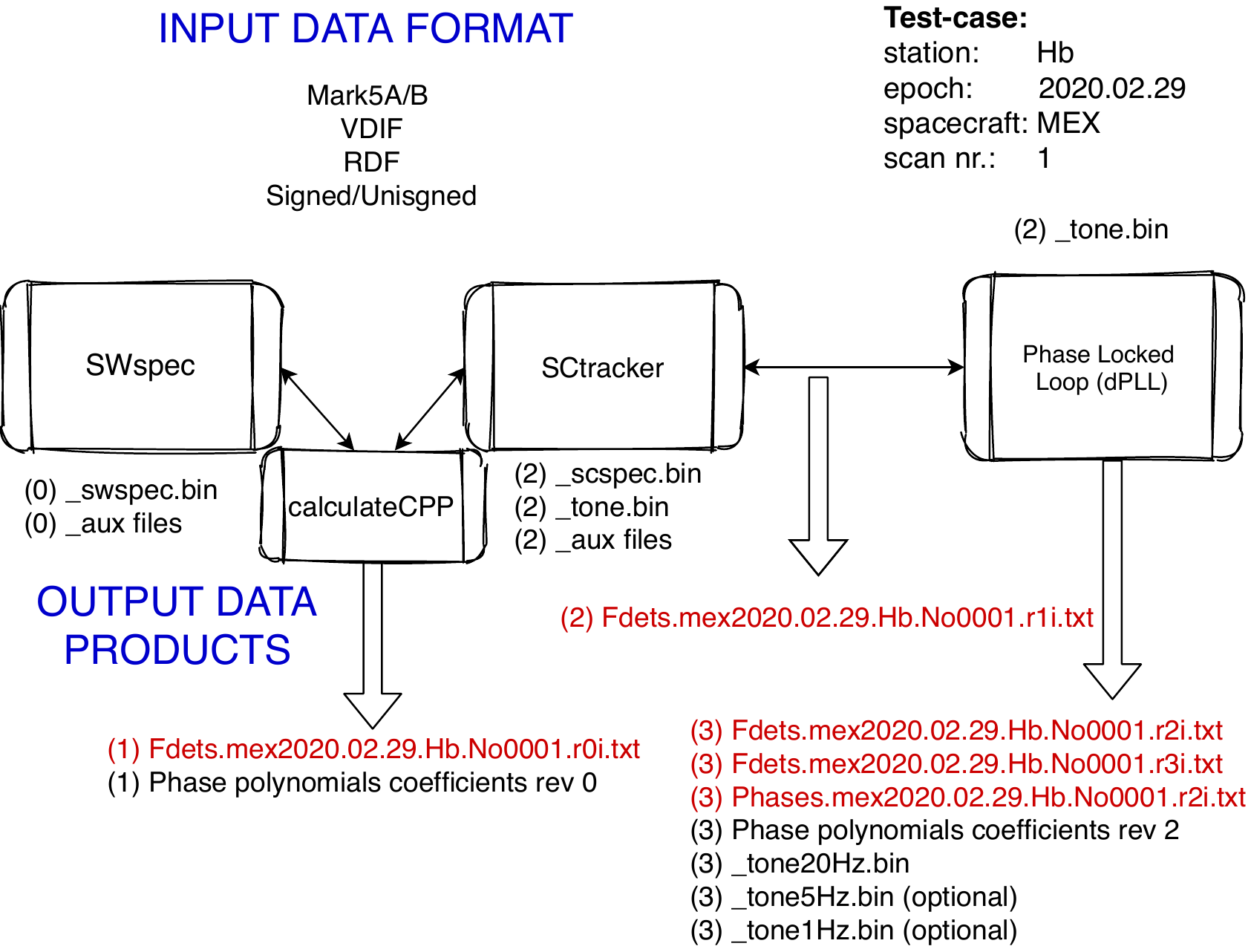}
    \caption{Data products obtained by running all three packages in \sw{SDtracker}: \sw{SWspec}, \sw{SCtracker} and \sw{dPLL}. The labels are based on a session observed on $2020.02.29$ at Hobart.}
    \label{fig:data_products}
\end{figure}

The main results obtained with \sw{SDtracker} are the topocentric frequency detections of the spacecraft carrier signal, the reconstructed and the residual phase with respect to the station reference clock. An example of the output data files generated in a session conducted on the MEX spacecraft with the Hobart-$12$ (Hb) radio telescope (Australia) on $2020.02.29$ is shown in Figure~\ref{fig:data_products}.

\sw{calculateCpp.py} generates the first set of Doppler detection and the residual frequency. The file is labelled with the code \verb!r0i! since it is the first iteration. The file contains four header lines with metadata including sky frequency, station, epoch, and $6$ columns of values in ASCII format: Modified Julian Date (MJD), sample timestamp [s], SNR, raw spectral power of the carrier peak (normalised to 1), frequency of the carrier [Hz], and the stochastic noise of the polynomial fit [Hz]. The topocentric frequency detections for deep space probes have accuracy better than $0.2\,$Hz at this stage, using a frequency resolution of $5\,$Hz and $5\,$s integration time with an input bandwidth of $8\,$MHz. \sw{calculateCpp.py} can also be used to verify the quality of the phase-stop after \sw{SCtracker}. The generated topocentric frequency detections and polynomial coefficients are then labelled with the code \verb!r1i!. These are rarely used since it is easier to work with narrow band rather than wide band signals.

\sw{dPLL.py} generates the second iteration of the topocentric frequency detections \verb!r2i!. Here, the input bandwidth is $2\,$kHz and the integration time varies between $1$ and $10\,$s based on the motion of the spacecraft and SNR of the carrier. For example, we use $10\,$s for interplanetary plasma scintillation studies~\citet{Molera2014}, but $1\,$s in the radio occultation experiments and aerobraking experiments~\citet{Bocanegra2019}. These residuals frequency have generally an accuracy better than $5\,$mHz. Finally, the topocentric frequency detections \verb!r3i! are generated after extracting the residual phase of the main carrier tone and reconstructing the frequency from the phase of the signal. The difference between the \verb!r2i! and \verb!r3i! revisions is small ($\leq 1\,$mHz), but noticeable in experiments where ultra-high accuracy is required~\citet{Litvinov2016,Litvinov2018}.

The residual phases file includes two columns with the timestamp and residual phase in radians for the whole scan. This file also contains four header lines summarising the epoch, station name, MJD, initial time, averaged SNR, and residual frequency after the second polynomial fit. The total number of samples (TS) is twice the output bandwidth times the scan length (TS=$2\cdot$BW$_{\rm o}\cdot T_{\rm span}$).

\section{Software features and performance}
\label{featperf}

\sw{SDtracker} is a compact software that can be easily deployed and utilised at any VLBI radio telescope, while offering a number of unique features to users interested in deep space tracking and spectroscopy. The software, initially released in $2010$, has undergone several upgrades during the past years. Many of these upgrades have focused on optimisation and computational performance, while others have added compatibility to the latest VLBI data formats. Some of these features and the current computational benchmarks are presented below.

\subsection{Software features}
\label{features}

The \sw{SDtracker} has been used over a thousand spacecraft tracking sessions to optimise its features and develop a robust software package.

\subsubsection{Multi-tone extraction}
\label{dor}

Spacecraft transponders often transmit specially formed harmonic subcarrier signals in order to support differential one-way ranging (DOR) tones. These sub-ranging tones are separated by a few kilohertz or megahertz from the main carrier signal. DSN DOR measurements use these generated tones to estimate the differential group delay and reduce the uncertainty of the measured delay due to instrumental and propagation effects~\citet{ccsds2019}.

\sw{SCtracker} was designed to extract separate measurements of the spacecraft carrier signal and its sub-ranging tones. The number of extracted tones is virtually unlimited, since these do not add computational load as the phase rotation is applied to the entire channel band at once. The frequency span can be known a priori from the spacecraft specifications or from a visual inspection of the first generated spectrum. The position of the tones with respect to the main carrier is specified in the tones offset file.

Each output tone contains time samples of the sub-ranging tones with the user-defined cut-off frequency. The output bandwidth is the same as the one used for the carrier signal. The ranging tones can be modulated signals and, at present, \sw{dPLL} does not attempt to demodulate them. Each tone file could still be processed individually in \sw{dPLL} to extract frequencies and phases.

\subsubsection{Radio Frequency Interference (RFI)}

RFI is a severe problem in radio astronomy and spacecraft tracking applications are also affected from it. It can be only a minor concern while working with spacecraft equipped with high-gain antennas, as the received signal is strong. However, it may become problematic in space missions at large distances from Earth or when equipped with low-gain antennas. Identifying the signal that belongs to the spacecraft is achieved with the \verb!Python! scripts in \sw{pysctrack}. When the spacecraft tone is strong, the scripts automatically identifies the signal and estimates the intrinsic Doppler shift. In the case of a weak carrier signal, or if severe interference is present (especially at the lower operational band around $2\,$GHz, the so called S-band), additional visual inspection is required by the user. We distinguish two types of RFI:

\begin{itemize}
\item \textit{Spurious random interferences} are present during short duration of a scan. Frequency peaks are randomly encountered throughout the spectra, but only for short periods of time. This type of interference is generally not problematic as the software automatically discards it. Its origin is usually associated with the instrumentation noise and RFI reflections.
\item \textit{Steady frequency tones} present constantly for the entire duration of the scan. The origin of these tones is associated with human-made signals, such as near-by tracking stations, communication towers, radar systems, etc. They could be confused with a spacecraft signal but can be often recognised since their frequency does not shift in time. A similar to an RFI impact on the signal detection can be produced by the phase calibration tones used at VLBI telescopes (see Section~\ref{pcal}).
\end{itemize}

We have developed a utility that autonomously distinguishes the RFI peaks from any carrier signal transmitted by satellites and spacecraft. The \sw{huntSC.py} script can also be found in the \sw{pysctrack} package. The program inspects the entire frequency bandwidth observed for potential RFI peaks. Based on SNR and the Doppler shift, it identifies all spacecraft tones present in the full bandwidth. The spacecraft carrier frequency is a priori known and varies on the order of hundred kHz depending on the spacecraft's motion. While the search for the tone is not usually needed, it is useful in multiple scenarios: for environments where multiple spacecraft communicate simultaneously with ground-stations (see Figure~\ref{fig:full_spectrum} for example); for the first time observing a particular spacecraft; for spacecraft with weak transmission power, or for locations with severe RFI.

\subsubsection{Cross-polarisation}

\sw{SWspec} can compute the cross-polarisation product of two data inputs. For example, for two data streams of different polarisation senses (i.e. left-hand and right-hand circular polarisations), \sw{SWspec} calculates the auto-correlation of each data stream and their cross-correlation product. There is an option in the control file to select two different input channels and their output file names. The main motivation to include cross-polarisation was to support observations of astronomical masers.

The current VLBI instrumentation follows a trend towards linearly polarised receivers. The VLBI Global Observing System (VGOS) community has developed a new broadband receiver ($2$--$14$ GHz) with linear horizontal and vertical feeds~\citet{Niell2018}. \sw{SWspec} can compute simultaneously both polarisations and then the cross-polarisation product. At present, to calculate the topocentric frequency detections and residual phase we use only one of the polarisations, usually the one with a higher SNR after \sw{SWspec}. Deep space craft usually transmit in right-hand circular polarisation. In the case of orbiter-lander communications, the polarisation may be linear thus their transmission is observed in both right- and left-hand circulars.

\subsubsection{Phase calibration tones}
\label{pcal}

The phase calibration (\textit{pcal}) tones consist of series of low-level monochromatic signals distributed across the frequency band. The phase calibration method is widely used in geodetic and astronomical VLBI for compensation of the instrumental phase errors generated by the station instrumentation. They appear as strong peaks in the spectral domain, despite representing around $1\,\%$ of the total power in the VLBI broadband signal.

\sw{SWspec} can extract the phase and amplitude of the \textit{pcal} tones in the observing band. A deeper analysis of the \textit{pcal} tones can also be performed (a)~by  filtering the signal with \sw{SCtracker}, as done for a spacecraft carrier, but without applying Doppler compensation, (b)~by analysing the filtered tone with \sw{dPLL.py}. The use of \textit{pcal} tones provide a direct measurement of the instrumental system noise at the telescope during spacecraft observations. \sw{SWspec} determines the reliability of the signal chain of a VLBI radio telescope and the diagnostics of the system performance in almost real-time as discussed by~\citet{Uunila2014}.
 
\subsection{Software performance}
\label{performance}

\begin{figure*}[!ht]
  \begin{tabular}{c c}
   \includegraphics[width=0.47\textwidth]{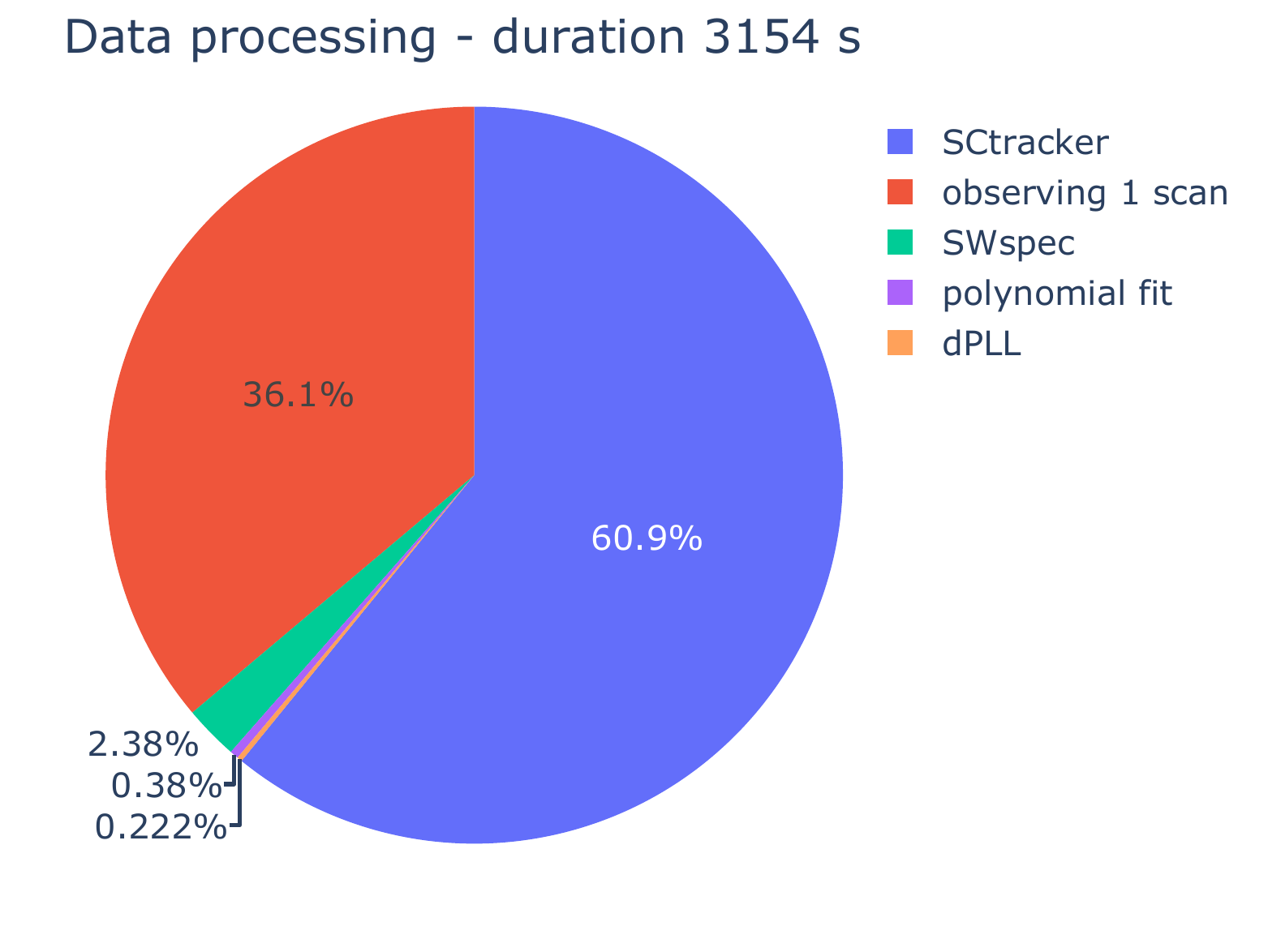} & \includegraphics[width=0.47\textwidth]{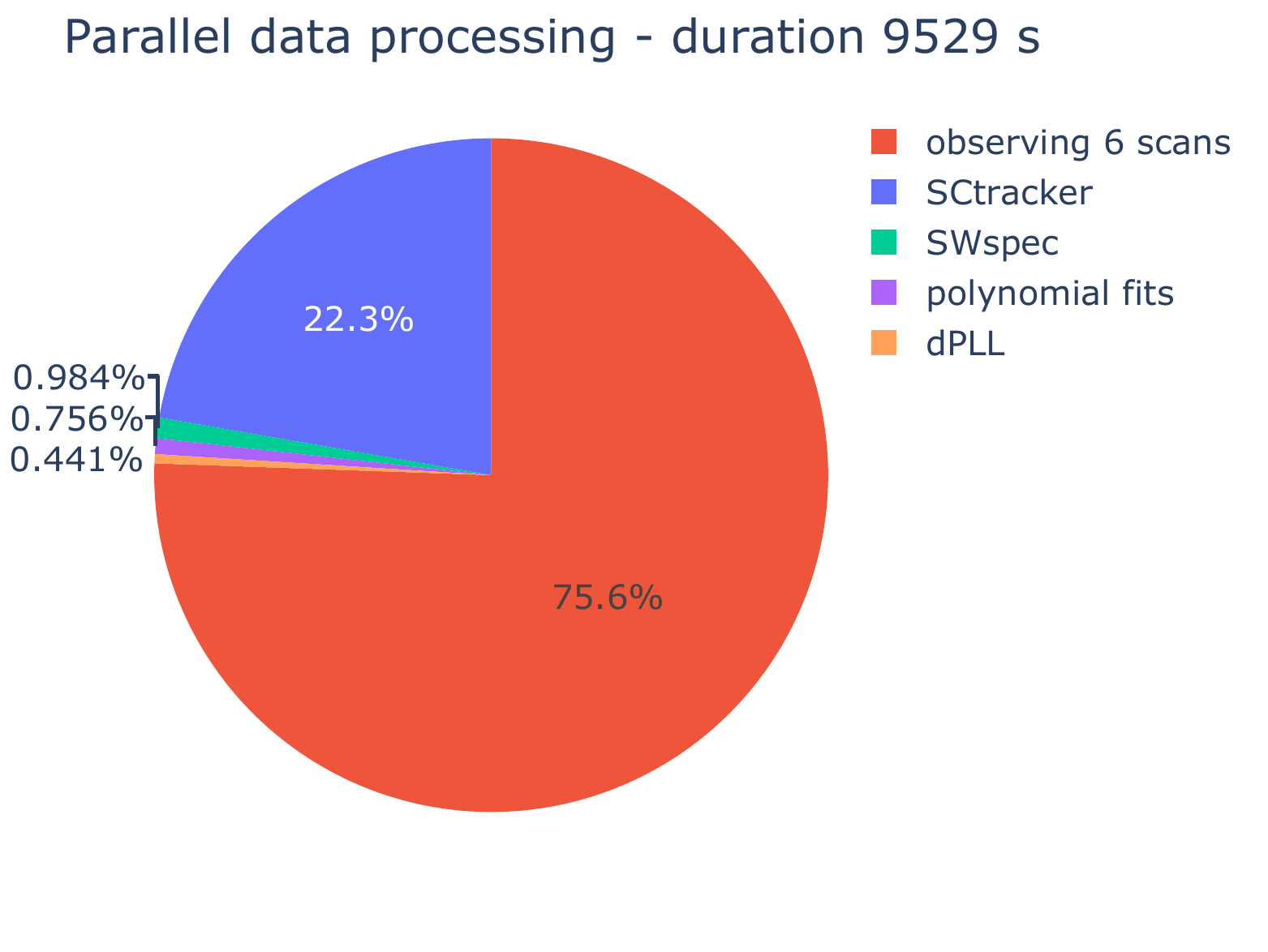} \\
   (i) & (ii) \\
  \end{tabular}
  \caption{(i) Processing time for the full pipeline of a single $19$-min scan. The configuration used here is $8\,$MHz bandwidth and $3.2\,$M FFT and $5\,$s integration time. (ii) Processing time for the full pipeline of six scans processed in parallel using the same observation settings.}
 \label{fig:pie}  
\end{figure*}

\sw{SWspec} and \sw{SCtracker} are compact applications intended for deployment on the generic setup of VLBI stations. They have been tested in most common Linux operative systems and macOS. The software takes advantage of the Intel Integrated Performance Primitives (IPP) libraries~\footnote{\url{https://software.intel.com/}; accessed on 2021.04.26} to optimise the code and maximise the performance of computationally-intensive tasks. Benchmarks on Linux platforms showed an improvement of $40\,\%$ in CPU performance using the IPP libraries compared to the standard Linux Fastest Fourier Transform in the West package. The software can be used on the same data acquisition system where the data have been recorded, avoiding large data transfers to the processing centre.

The full data processing pipeline for a typical $19$-min tracking scan takes around $50$-$55$ minutes including the observing time. The breakdown of the different tasks is shown in Figure~\ref{fig:pie} (left panel). In this example, we used \sw{SWspec} with four-cores and \sw{SCtracker} in single-core mode. In $2010$, the processing of a single $19$-min scan took about two hours of computational time~\citet{Molera2012}. The latency has been improved by more than $200\,$\% between $2010$ and $2020$. The right panel shows the total computational time when processing six $19$-min scans. Since the data processing is performed in parallel, the largest component to the overall time is now the antenna observing time. Six scan can be computed in parallel within $160$ minutes.

\sw{SCtracker} was developed to run on a single-core processors. The parallelisation is done via a \verb!Python! script included in \sw{pysctrack}, which processes multiple scans simultaneously. \sw{SWspec}, on the other hand, is capable of processing a single scan on a multi-core system; this can be selected from the configuration file. A benchmark of the \sw{SWspec} performance with different input and multi-core settings is presented in Table~\ref{tab:performance}. I/O operations become the main bottleneck when reading the file from a single hard disk and using more than six cores in parallel. In that case, it is recommended to store the files in a RAID system or a vbs-type\footnote{VLBI FS:~\url{github.com/jive-vlbi/}; accessed on 2021.05.01} file disk system.

 \begin{table}[!htb]
 \centering
 \caption{\sw{SWspec} performance benchmarks with different input settings and using parallel vs single processing modes. The table includes: integration time (dT), number of FFT's points (FFT's) [million points], bandwidth (BW) [MHz], FFT overlap (Ovl), number of cores in parallel ($\#$c), FFT calculation time (FFT T), total time (Total T).}
 \label{tab:performance}
 \small
 \begin{tabular}{c|c|c|c|c|c|c}
  \hline\hline
  dT &  FFT's & BW & Ovl & $\#$c & FFT T & Total T \\
 \noalign{\smallskip}\hline\noalign{\smallskip}
  1 s & 3.2 M & 16 & 1 & 1 & 185 s & 200 s \\
  2 s & 3.2 M & 16 & 1 & 1 & 171 s & 202 s \\
  5 s & 3.2 M & 16 & 1 & 1 & 178 s & 197 s \\
  1 s & 3.2 M & 16 & 2 & 1 & 515 s & 530 s \\
  2 s & 3.2 M & 16 & 2 & 1 & 569 s & 560 s \\
  5 s & 3.2 M & 16 & 2 & 1 & 423 s & 465 s \\
  5 s & 3.2 M & 16 & 2 & 2 & 241 s & 275 s \\
  5 s & 3.2 M & 16 & 2 & 4 & 191 s & 222 s \\ 
  5 s & 3.2 M & 16 & 2 & 6 & 174 s & 210 s \\
  5 s & 3.2 M & 16 & 2 & 8 & 155 s & 187 s \\
  2 s & 6.4 M & 16 & 2 & 2 & 432 s & 480 s \\
  5 s & 6.4 M & 16 & 2 & 2 & 255 s & 299 s \\
  2 s & 6.4 M & 16 & 2 & 8 & 164 s & 213 s \\
  5 s & 6.4 M & 16 & 2 & 8 & 100 s & 155 s \\
  5 s & 1.6 M & 16 & 2 & 2 & 190 s & 209 s \\
  5 s & 6.4 M & 16 & 2 & 8 & 142 s & 169 s \\
  5 s & 1.6 M & 16 & 1 & 8 &  82 s & 120 s \\
  5 s & 3.2 M & 16 & 1 & 8 & 106 s & 141 s \\
  2 s & 6.4 M & 32 & 2 & 2 & 961 s & 993 s \\
  5 s & 6.4 M & 32 & 2 & 8 & 461 s & 472 s \\
  2 s & 3.2 M &  8 & 2 & 2 & 135 s & 140 s \\
  5 s & 3.2 M &  8 & 2 & 8 &  76 s & 89 s \\ 
  \hline\hline
 \end{tabular} 
\end{table}

\begin{figure*}[ht!]
 \includegraphics[width=1.0\textwidth]{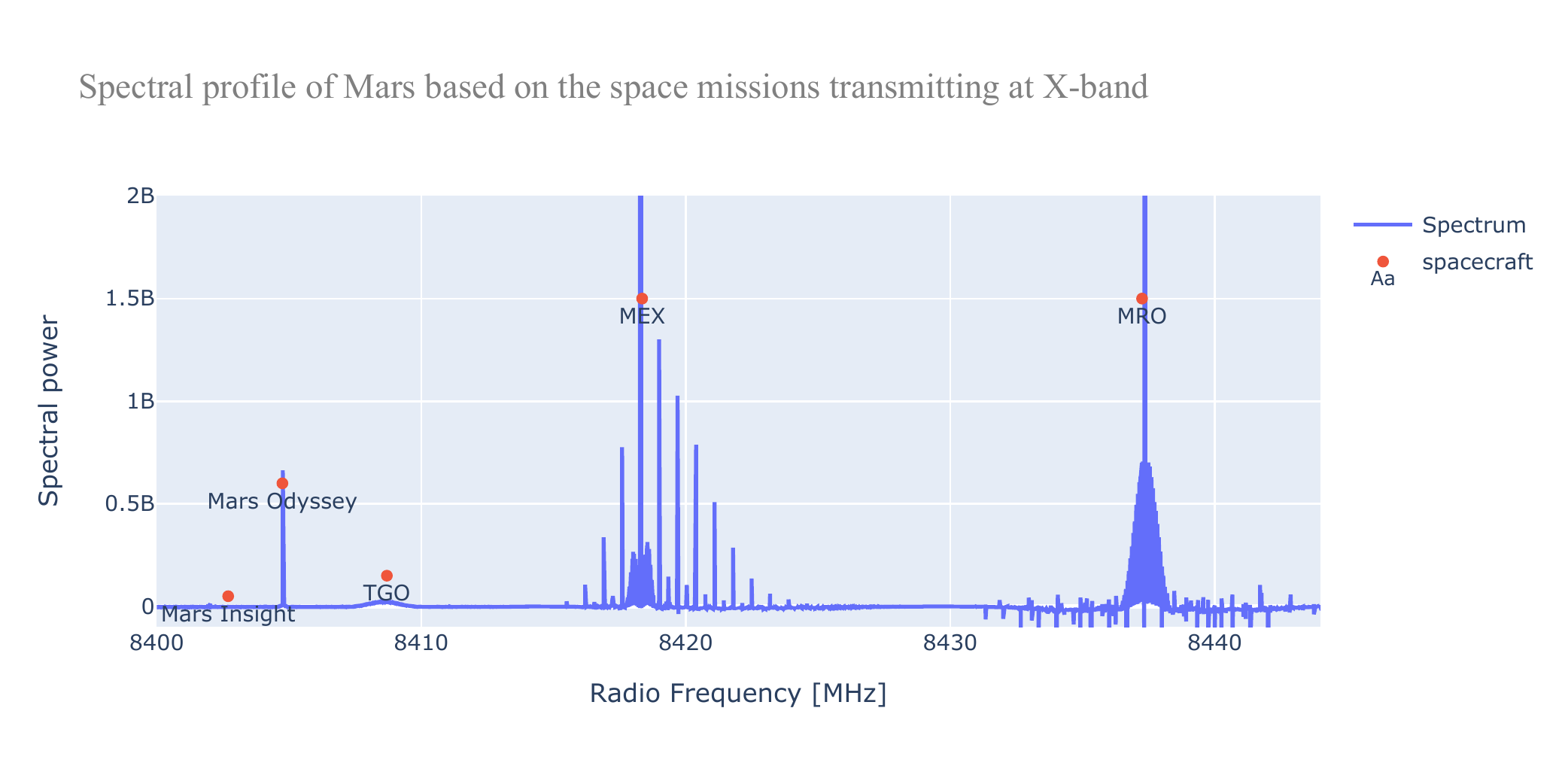}
 \caption{Radio spectrum at X-band showing signals from several spacecraft operating on the surface of Mars (Mars Insight) and in various aerocentric orbits (Mars Express, Mars Odyssey, Mars Reconnaissance Orbiter, and ExoMars Trace Gas Orbiter).}
 \label{fig:full_spectrum}
\end{figure*}

VLBI generates large volumes of raw data, with typical data rates of several gigabits per second per station resulting in a total volume of accumulated raw data per experiment of hundreds of terabytes. In order to maximise the sensitivity of VLBI observations, observing systems are widening the frequency bandwidth. Current receivers are capable of collecting radio signals continuously from $2$ to $14\,$GHz. This observed bandwidth is then segmented into frequency bands, or frequency channels, which are then recorded by DAS~\citet{Molera2012}.

Several DAS types are utilised in modern VLBI facilities. The most widely used are the Flexbuff systems developed at the Mets\"{a}hovi Radio Observatory and the Mark$5$/$6$~\citet[see review in][]{Lindqvist2014}. These units record multiple digitised frequency channels in files ranging in size from tens to hundreds of gigabytes. The data format includes the observing mode, antenna code, and the time of the observations, among other metadata. The spacecraft tracking software is compatible with most of the major versions of the data format developed for VLBI in the recent years, including Mark$5$A/B, RadioAstron Data Format (RDF) and the latest VLBI Data Interchange Format (VDIF). \sw{SWspec} and \sw{SCtracker} can also read non-formatted data without timestamps produced by non-standard radio astronomical instrumentation.

\section{Software validation}
\label{validation}

The \sw{SDtracker} software has been validated in various spacecraft tracking applications presented in this section with corresponding data processing steps.

\subsection{Data processing pipeline}
\label{analysis}

Radio telescopes equipped with VLBI instrumentation provide for acquisition of wide–band data, making possible to capture multiple spacecraft radio signals simultaneously. The transmission band for deep space navigation at X-band, between $8400$ and $8450\,$MHz, fits in one or two standard VLBI frequency channels. In the case of Martian missions, where no less than $10$ craft are operating in $2021$, all spacecraft carrier signals can be detected and processed simultaneously by a single VLBI radio telescope. Figure~\ref{fig:full_spectrum} shows the data recorded in one experiment in February $2020$ in which five craft (one lander and four orbiters of Mars) were transmitting simultaneously to ground tracking stations. We can distinguish NASA Mars Insight ($8404.5\,$MHz), NASA Mars Odyssey ($8407.2\,$MHz), ESA ExoMars Trace Gas Orbiter ($8409.7\,$MHz), ESA Mars Express ($8419.9\,$MHz) and NASA Mars Reconnaissance Orbiter ($8439.7\,$MHz). The power received, the frequency and the stability of the phase are different and they depend on the characteristics of each spacecraft. The topocentric frequency detections in a session can vary by an order of few mHz to several Hz per second due to the Doppler shift.

\begin{figure*}[htbp]
  \begin{tabular}{c c}
   \includegraphics[width=0.45\textwidth]{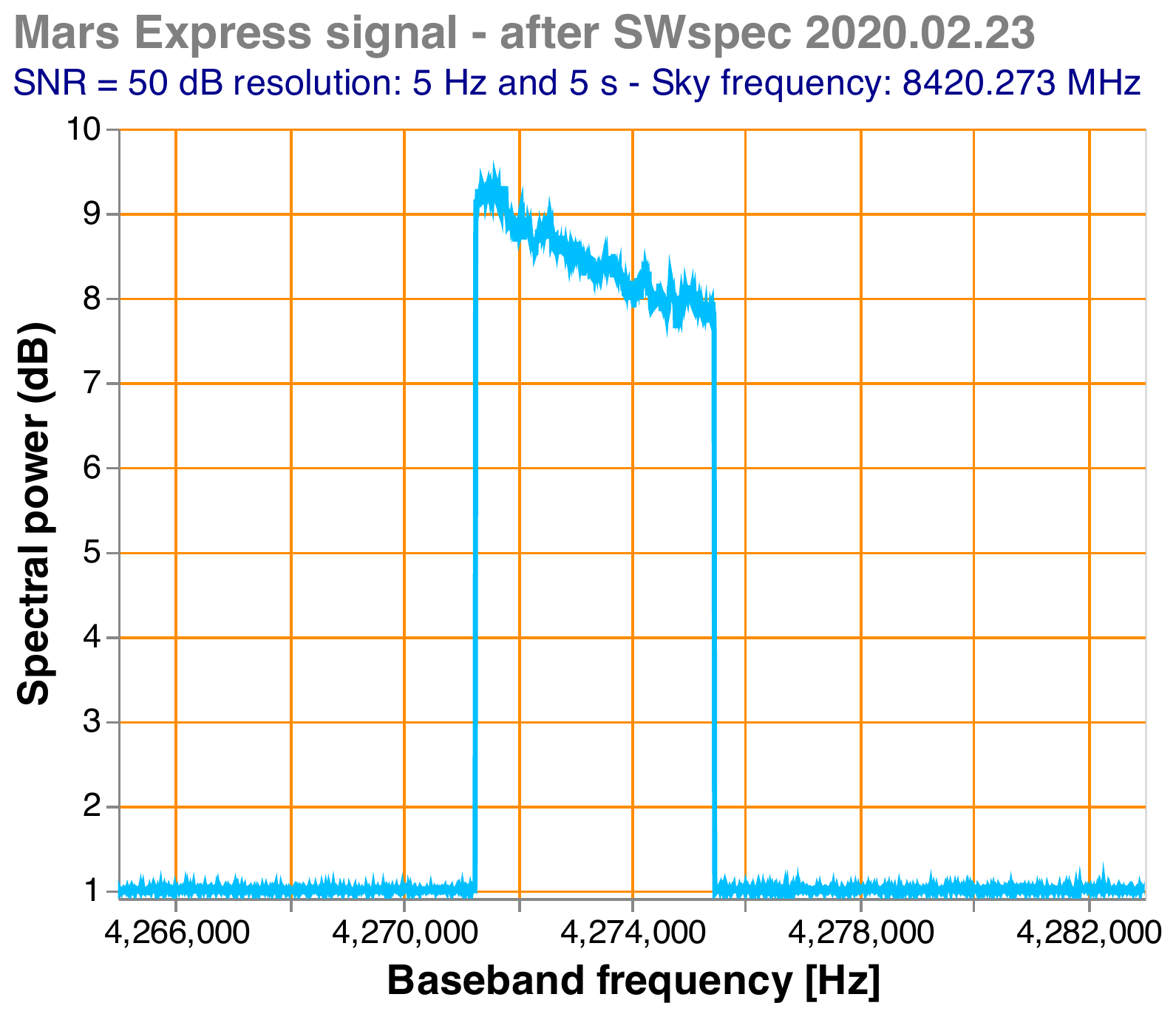} &
   \includegraphics[width=0.45\textwidth]{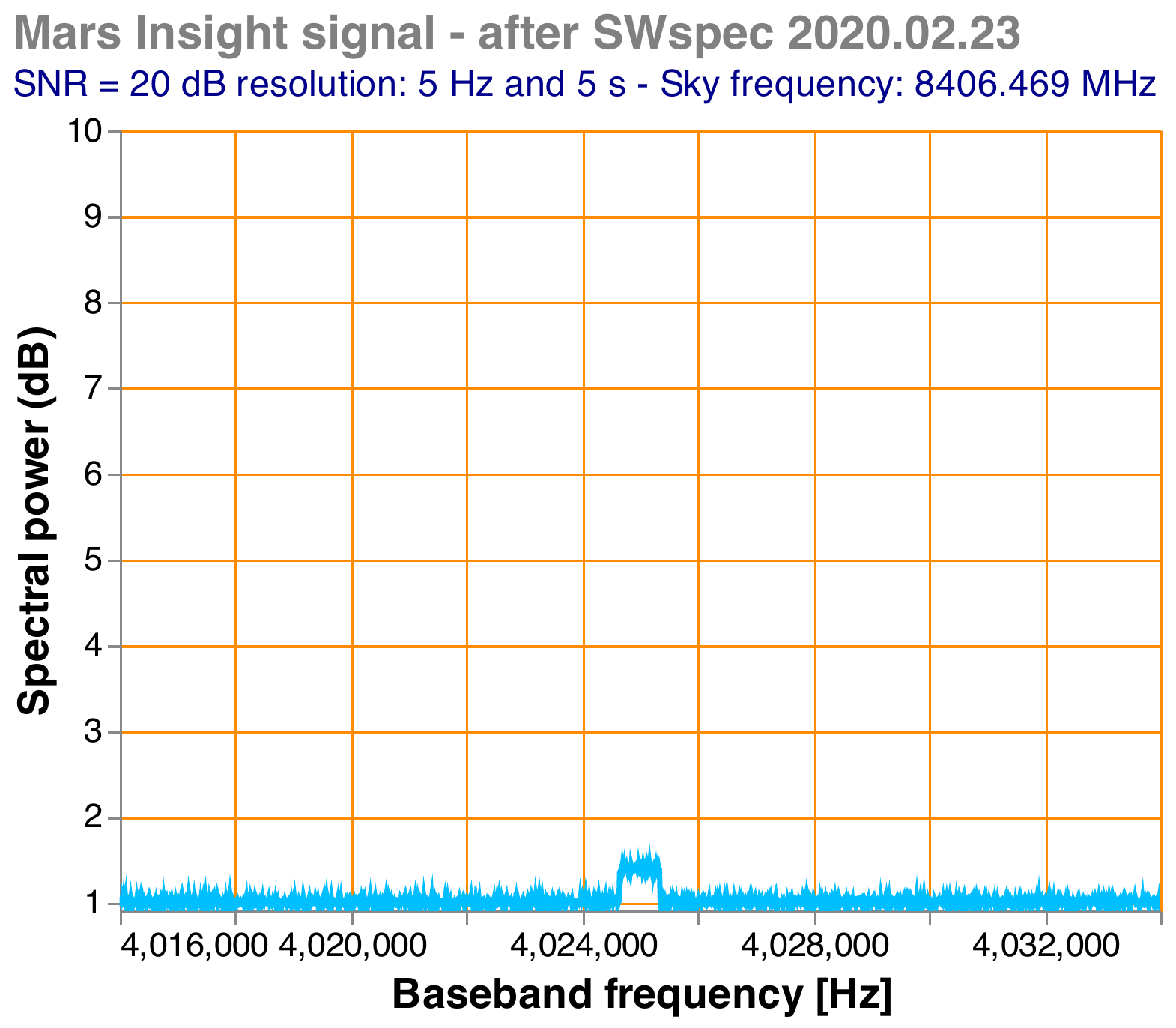} \\
   (i) & (ii) \\
   \includegraphics[width=0.45\textwidth]{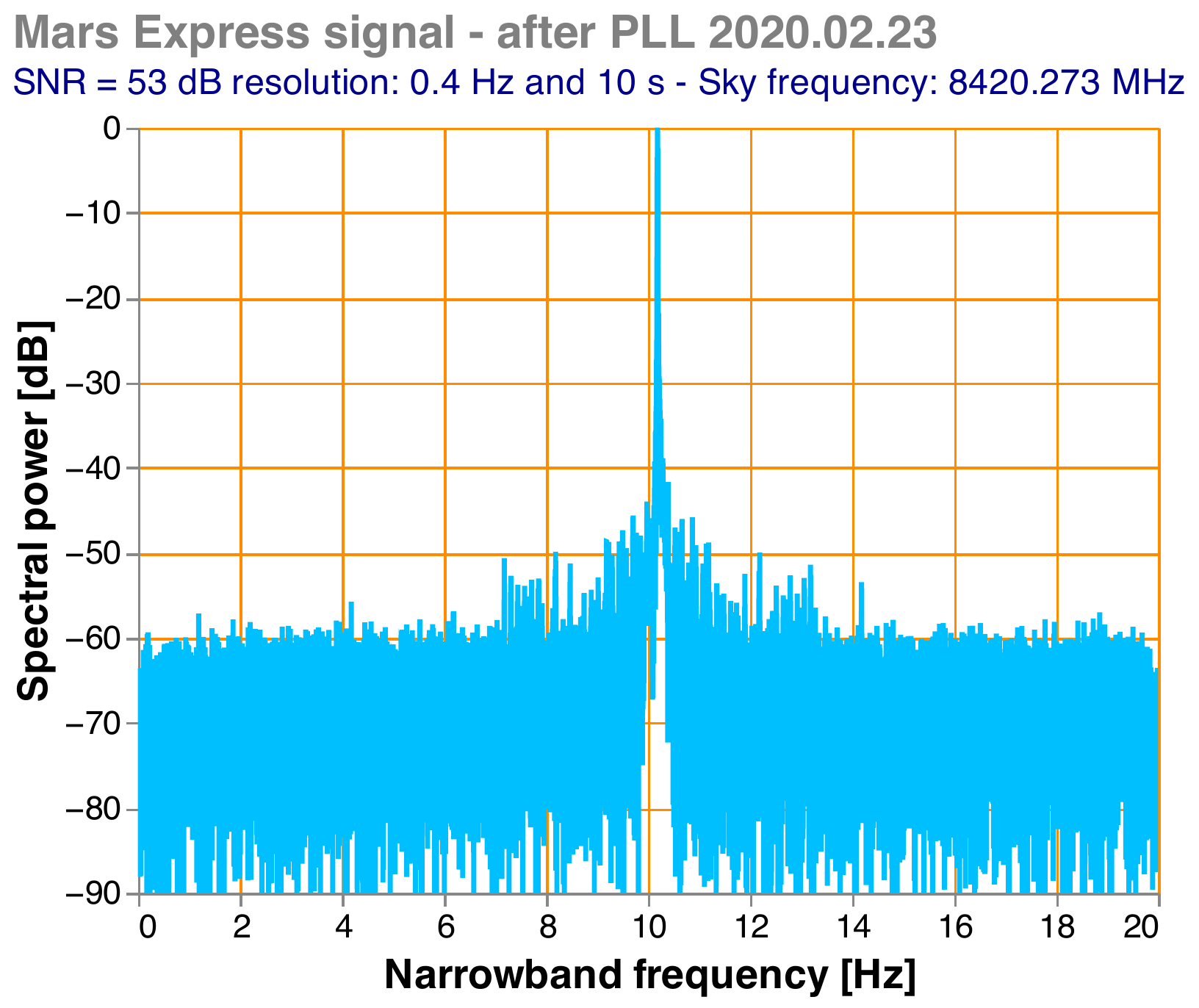} &
   \includegraphics[width=0.45\textwidth]{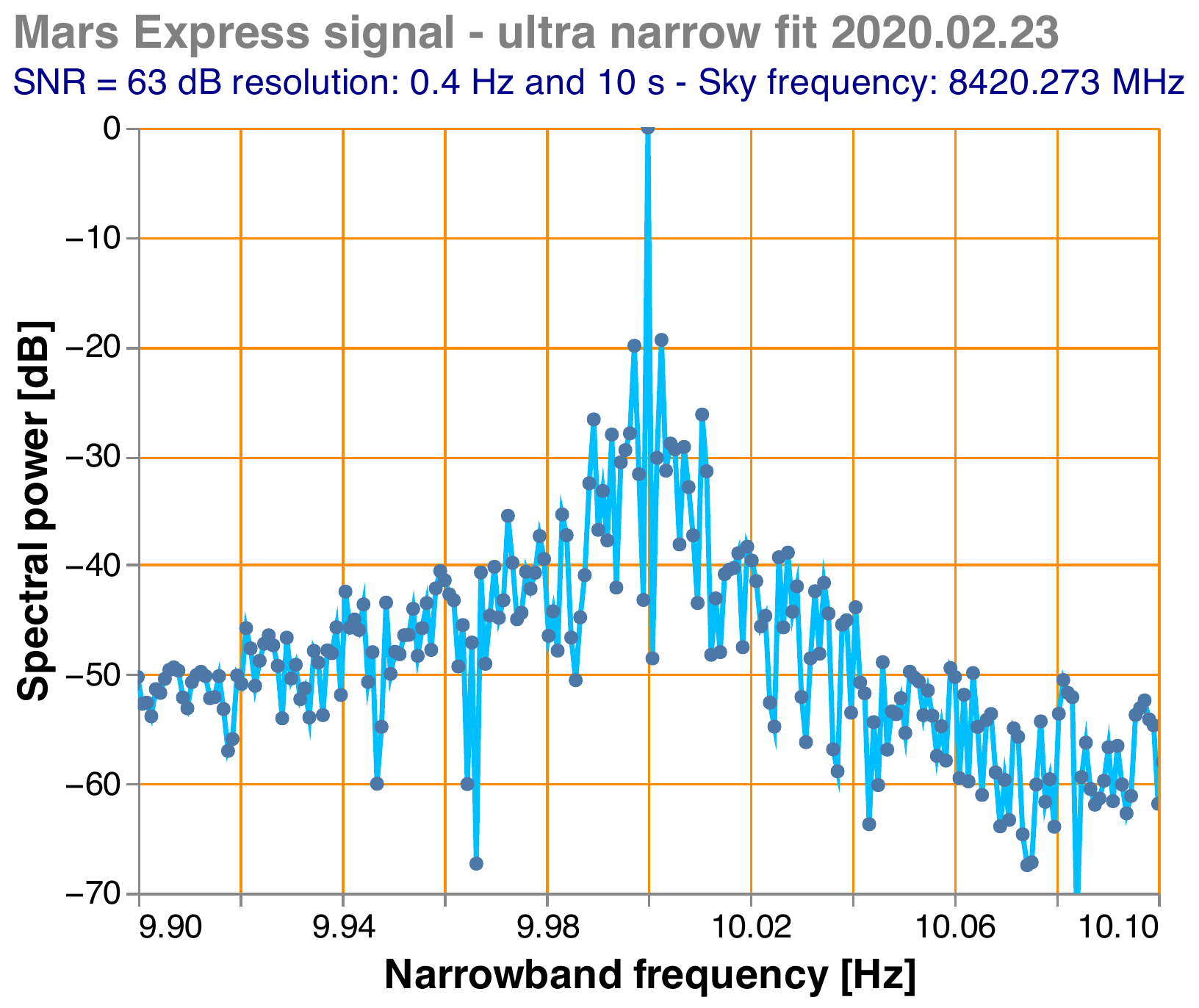} \\
   (iii) & (iv) \\
  \end{tabular}
  \caption{(i) Spectral power of Mars Express observed with the 30-m Ceduna radio telescope on $2020.02.23$. The plot shows the Doppler shift on a $19$-min scan using dF of 5 Hz and $5\,$s integration time. (ii) Spectral power of Mars Insight observed with the $65$-m Tianma radio telescope on the same epoch. The SNR of Mars Insight was $20.1\,$dB and a Doppler shift of $400\,$Hz in $19$-min (iii) MEX carrier tone detected in narrow band ($20\,$Hz) after $1140\,$s of integration with a frequency resolution of $0.4\,$Hz. The SNR of the signal is $51\,$dB and the tracking of the signal is within a precision of $0.8\,$mHz. (iv) Same carrier tone detected in ultra narrow band ($1\,$Hz), now the SNR improves to almost $60$ dB.}
 \label{fig:sctone}  
\end{figure*}

This Figure~\ref{fig:full_spectrum} illustrates the X-band electromagnetic spectrum captured at a VLBI radio telescope. We can run the full software pipeline to any detected planetary spacecraft to extract the intrinsic parameters of the carrier signals. Multiple stages of the processing pipeline are presented in the four panels of Figure~\ref{fig:sctone}. Panel (i) shows the results of \sw{SWspec}, that is the accumulated $227$ integrated spectra ($5\,$s) for a scan of $19$ minutes of the Mars Express carrier signal acquired with the $30$-m Ceduna radio telescope on $2020.02.23$. The plot shows the Doppler shift range of $\sim 4.3\,$kHz in $19$-min ($3.7\,$Hz/s or $134\,$mm/s) and the peak maximum of the normalised spectral (SNR $\sim40\,$dB).

For comparison, panel (ii) shows the signal of Mars Insight after \sw{SWspec} for the same scan. Instead of using data from Ceduna, the plot presents data from Tianma (65-m) because the lander signal was weak. NASA's Mars Insight is a lander deployed on the surface of Mars and equipped with a low-gain antenna. Due to the lower transmission power combined with a season of dust storms on Mars, the detection was $20$--$30\,$dB weaker than other standard missions orbiting Mars at the moment. The Doppler shift was also small, around $400\,$Hz over $19$ minutes ($0.35\,$Hz/s). The average SNR in the scan was only $20.1\,$dB, while the SNR for Ceduna using $10$-s integration time was $11\,$dB (i.e. barely detectable).

The detection of the initial tone is improved by using all three packages in \sw{SDtracker} to completely phase-stop the spacecraft carrier in a narrow band. In panel (i), the spectral tone is spread across multiple frequency channels due to the Doppler shift introduced by the motions of the ground-station and spacecraft. In panel (iii), the spacecraft carrier signal is corrected for the Doppler shift with a mHz resolution. The phase correction includes the compensation applied firstly in \sw{SCtracker} and then in \sw{dPLL}.

Finally, the panel (iv) shows the result of applying all the possible corrections with the third set of phase polynomials to the carrier tone in a ultra-narrow band of a single Hz. In this case, we can see that all spectral power is distributed in a single spectral bin, for the duration of the entire scan. Demonstrating that the phase-compensation has worked as expected to stop the signal. The SNR now was almost $60$ dB.

\begin{figure}[ht!]
   \includegraphics[width=1.0\columnwidth]{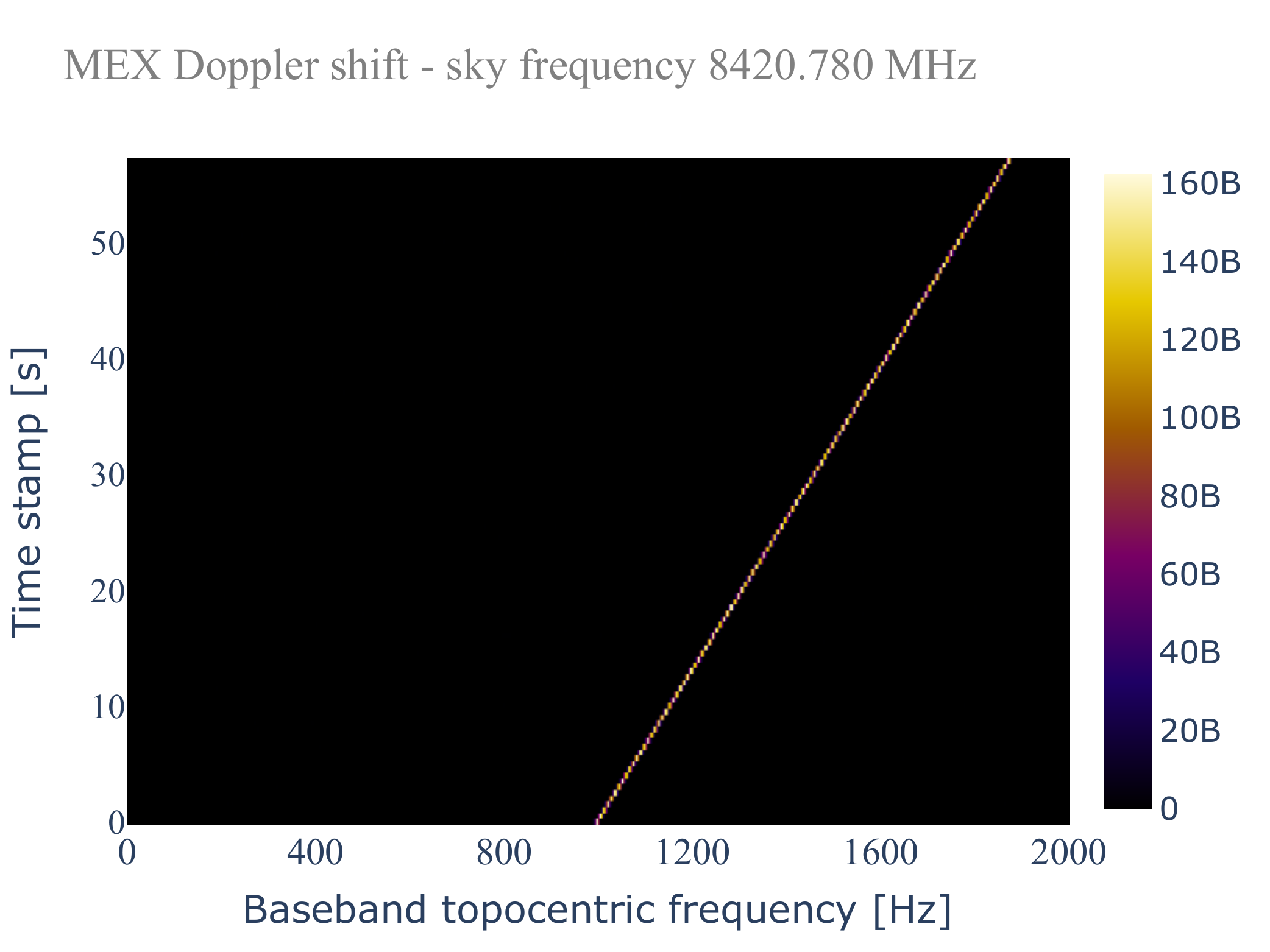}
  \caption{Dynamic spectrum of the MEX Doppler shift observed by the Ceduna radio telescope on $2020.06.21$, with a topocentric frequency shift rate of $15\,$Hz/s.}
 \label{fig:fdets}
\end{figure}

\begin{figure}[ht!]
   \includegraphics[width=1.0\columnwidth]{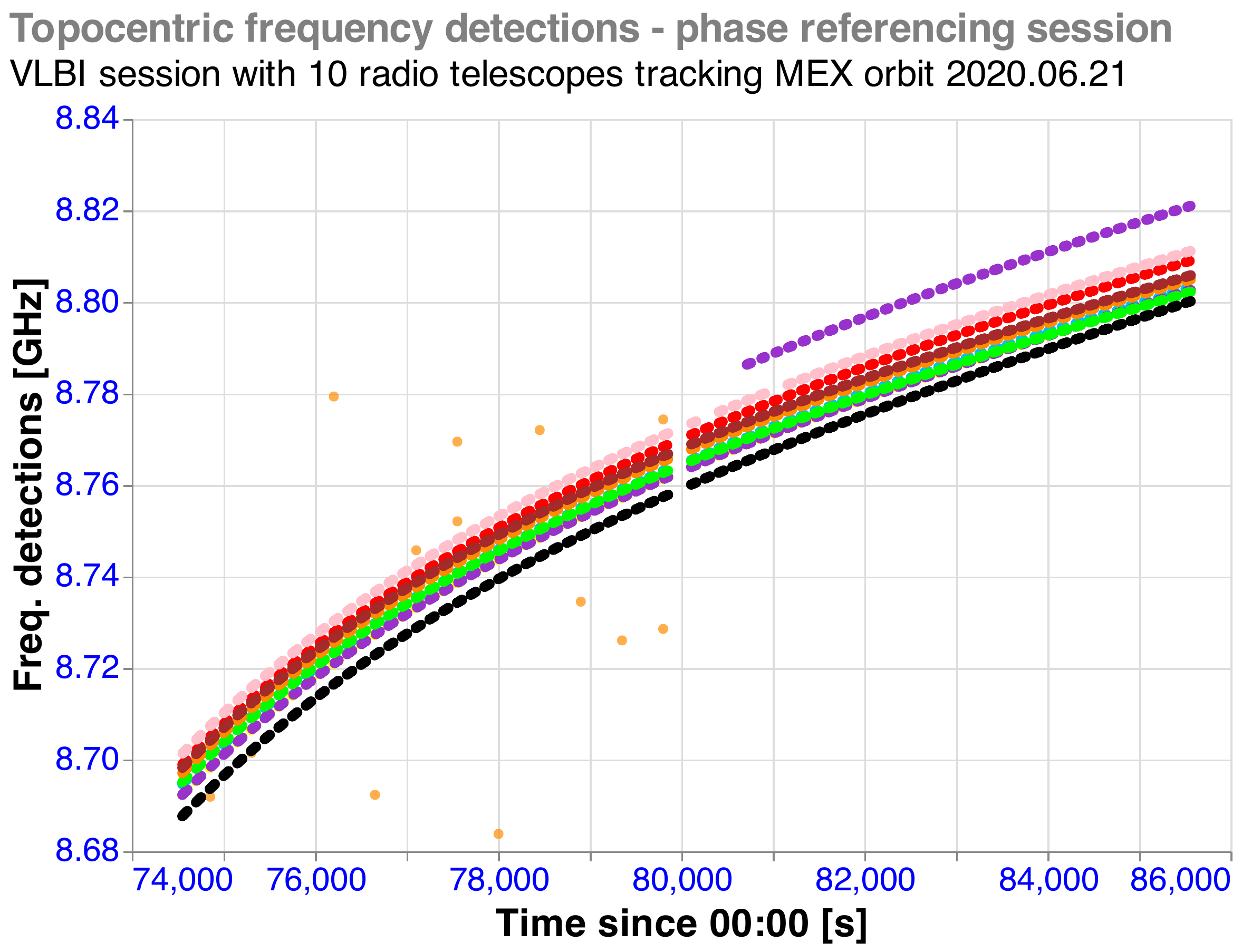} 
  \caption{Apparent topocentric frequency detections observed with multiple radio telescopes on a MultiView VLBI phase referencing session. The scans were $1$-min long and alternated with reference sources.}
 \label{fig:fdets2}
\end{figure}

After \sw{SWspec}, the first estimate of the apparent topocentric frequency detections (\verb!r0i!) of each spacecraft signal are calculated. The Doppler shift depends on the spacecraft tracked, especially on their orbits, and on the ground-station on Earth. Examples of apparent topocentric frequency detections are shown in Figure~\ref{fig:fdets} and~\ref{fig:fdets2}. In the first plot, a dynamic spectrum of the MEX signal shows the total Doppler shift ($874\,$Hz) during a $1$-min scan as the spacecraft moved away from the line of sight of the receiver. The Doppler shift rate was $15\,$Hz/s. The single scan was observed on $2020.06.21$ with Ceduna radio telescope. This dynamic range was generated from the apparent topocentric frequency detections calculated after \sw{dPLL} (\verb!r2i!)

In Figure~\ref{fig:fdets2} the topocentric frequency detections of multiple telescopes of the Mars Express MultiView phase referencing experiment are shown. The experiment was observed on $2020.06.21$ with $10$ VLBI radio telescopes. In this experiment, a fast nodding cycle of one minute on-spacecraft and one minute on each of the three reference sources was used. The experiment lasted three hours with the following antennae participating: Hobart-$26$, Hobart-$12$, Katherine, Yarragadee, and Ceduna (Australia), Warkworth (New Zealand), Tianma and Urumqi (China), Hartebeesthoek (South Africa) and Zelenchukskaya (Russia). The topocentric frequency detections are highly correlated, since the main contributor to the Doppler shift is the spacecraft motion. Differences are driven by the different motions of each station which depend on their terrestrial coordinates.

\begin{figure*}[ht!]
\centering
 \includegraphics[width=0.8\textwidth]{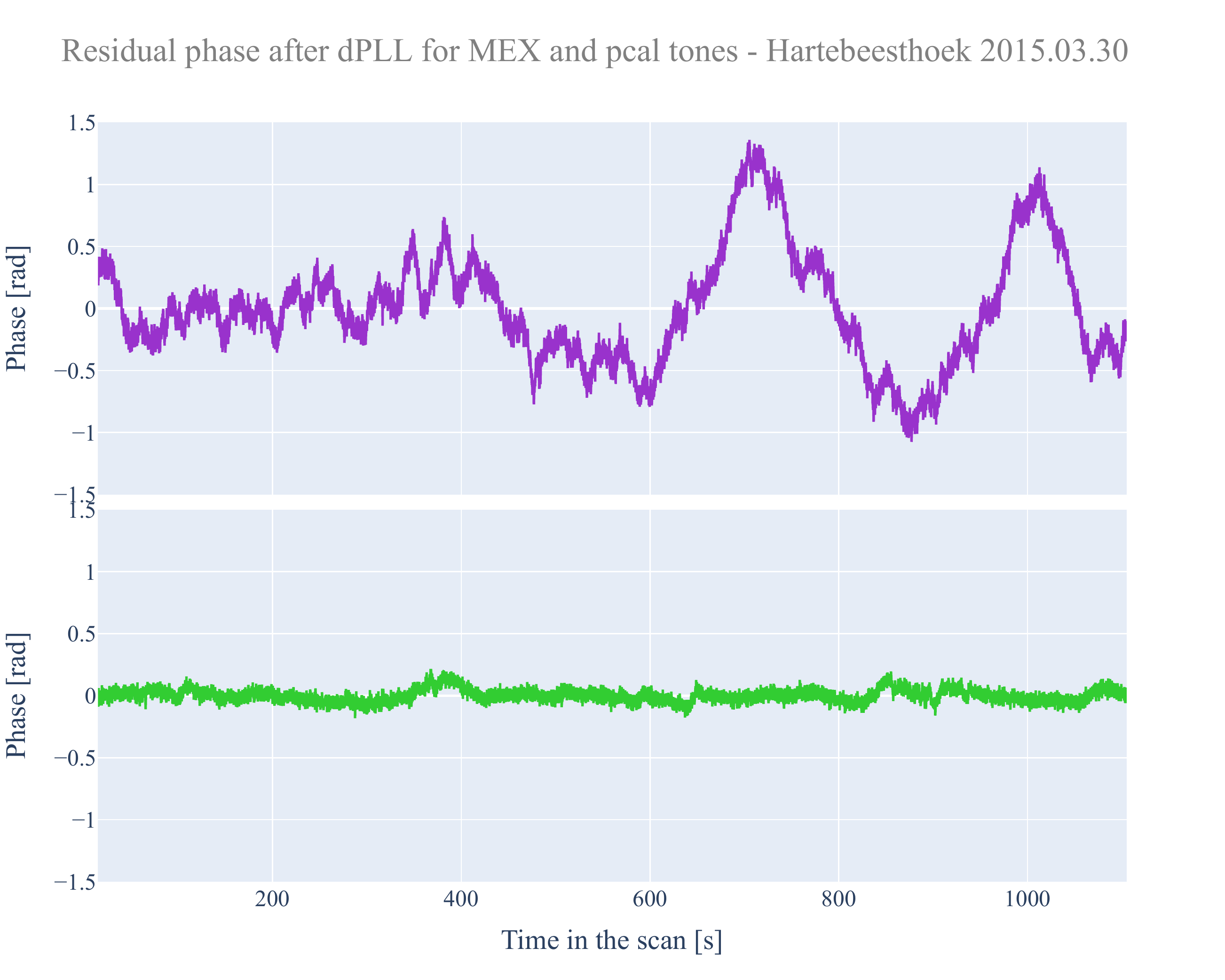}
 \caption{Phase extraction from the Mars Express signal (\textit{top}) and telescope phase calibration tone (\textit{lower panel}). The data were collected on $2015.03.30$ with the Hartebeesthoek $26$-m telescope in South Africa for interplanetary scintillation studies.}
 \label{fig:phases}
\end{figure*}

The main data products of the \sw{SDtracker} also include the spacecraft carrier phase and the residual phase with respect to local reference clock. The variations of the residual phase usually range from $100\,$mrad to $20\,$rad~\citet{Molera2014}. Figure~\ref{fig:phases} presents a comparison of the residual phase retrieved from a spacecraft carrier (upper panel) and from a \textit{pcal} tone (bottom panel) internally generated for the calibration of the system noise. Both tones were present in the same scan recorded at Hartebeesthoek on $2015.03.30$. MEX spacecraft was at $2.35\,$AU with a solar elongation angle of $19.5\,$degrees. The phase of the spacecraft signal fluctuates due to the propagation media between observer and target. The major contribution is due to the interplanetary plasma scintillation. For this particular experiment, the standard deviation ($\sigma_{\rm sc}$) was $0.521\,$rad. Observations at high solar elongations, at low horizon and/or during bad weather conditions, the phase fluctuations can be dominated by tropospheric and ionospheric turbulence. The \textit{pcal} tone provided an estimate of the system phase noise $\sigma_{\rm sn}$, which was equal to $0.014\,$rad.

Breaks in the residual phase can occur due to fast changes in the state vectors of the spacecraft, low SNR because of a low transmission power or a insensitive ground antenna, a data gap caused by a communications failure, and recording problems, –– among others. Some of these breaks can be fixed by tuning the time lags, spectral resolution, FFT zero-padding or overlapped factors in \sw{SCtracker} and \sw{dPLL}. Other breaks might require the recorded data to be split into multiple sub-scans and for the data processing to be performed on each of them individually. For example, this is the case for the gravitational redshift experiment carried out with the RadioAstron mission, in which the operations mode between spacecraft and ground-station can switch from one-way to two-way modes multiple times in a single scan~\citet{Litvinov2016,Litvinov2018}.

\sw{SDtracker} can process data obtained in an open-loop regime, which can be implemented as one-, two- or three-way mode. In one-way mode, the spacecraft transmits the signal to the ground station using an onboard ultra stable oscillator (USO) as the frequency reference. In the two-way mode, the ground station sends a signal to the spacecraft that is locked to a local frequency standard (typically a hydrogen maser) and the spacecraft retransmits the signal to the same ground station. In the  three-way mode the receiving and transmitting ground stations are different. The main difference between the one- and the two-/three-way regimes is that, in the latter, the phase stability is higher in most cases. Due to the better stability of the signal, some radio science experiments are preferably conducted in the two-/three-way mode. However,~\citet{Pogrebenko2004} demonstrated successful tracking of the Huygens descent in Titan's atmosphere supported by the Huygens onboard USO. Another example of a successful use of a one-way mode is the radio occultation experiment~\citet{Bocanegra2019}. In the latter case, the one-way mode makes possible retrieval of both ingress and egress occultation profiles, as opposed to only ingress in the two-/three-way mode.

\subsection{Tracking of deep space missions and Earth satellites}

The \sw{SDtracker} software has been used for different types of spacecraft operating at both interplanetary distances as well as Earth satellites at sufficiently high orbits.

\subsubsection{Deep space planetary missions}

The ESA's Venus Express (VEX) was the first spacecraft used for extensive tests of the tracking software presented here. VEX was launched in $2005$ and operated until December $2014$. Most of the communication operations were carried out at X-band ($8419\,$MHz), including some at S-band ($2296\,$MHz). We conducted a total of $319$ VEX sessions using VLBI radio telescopes.

Another extensively observed spacecraft has been the ESA's Mars Express (MEX), operational since $2003$. MEX sessions with VLBI radio telescopes are still being conducted at the time of this writing in $2021$. Furthermore, a number of other spacecraft have been tracked by our group in the framework of various science projects since $2009$. In addition to their specific science objectives, these observations enabled us to verify and update the software. In particular, we modified the software to maintain its compatibility with the progressing spacecraft communications and VLBI instrumentation. Table~\ref{tab:observations} lists all the observations conducted between $2009$ and $2021$ for which the software described in this paper has been used. The main carrier frequency is an approximation as this can vary by several tens and hundreds of kHz.

\begin{table*}
  \caption{List of spacecraft observed through $2021.03.31$ used in developing \sw{SDtracker} at X-band, including the spacecraft name, epochs of the sessions (1), approximate frequency of the main carrier (2), station code (3), mean SNR of the carrier (4), stochastic frequency noise (5), and Doppler shift range over a $19$-minute scan (6). These values depend on the epoch, ground-station and settings used in \sw{SWspec}, hence, we used a standard session.}
  \label{tab:observations}
  \begin{tabular}{l | c c c c c c}
  \hline\hline
    Spacecraft or vehicle & Period & Carrier freq & Station & SNR & Doppler noise & Doppler range \\
    & & [MHz] & Code & [as in Eq~\ref{eq:SNR}] & [mHz] & [kHz] \\
    \hline
    ESA BepiColombo       & 2020-21 & 8420.29 & Cd &  6063 &   84 &  0.69 \\
    ESA Mars Express      & 2010-21 & 8420.75 & Cd & 10000 &   30 &  20.0 \\
    ESA Rosetta           & 2016    & 8211.87 & Mh &   916 &   40 &  0.85 \\
    ESA Trace Gas Orbiter & 2018-20 & 8409.72 & Tm & 11500 &   90 &  3.51 \\
    ESA Venus Express     & 2009-14 & 8418.15 & Ht &  9500 &   26 &  1.50 \\
    \hline
    Roscosmos RadioAstron & 2011-15 & 8399.70 & Wz & 32000 &    3 &  1.10 \\
    \hline
    JAXA Akatsuki          & 2010    & 8431.29 & Ww &   150   &  -    &  -    \\
    JAXA Ikaros            & 2010    & 8410.92 & Ww &   100   &  -    &  -    \\
    \hline
    NASA Insight           & 2020    & 8404.50 & T6 &  120  &  300 &  0.40 \\
    NASA Juno              & 2020    & 8435.32 & Ho &  450  &  200 &  0.70 \\
    NASA MRO$^*$           & 2016-19 & 8439.71 & Cd & 8000 & $20000^*$ &    35.0 \\
    NASA Stereo A \& B     & 2011    & 8446.23 & Wz & 3000  & 200  &  0.45 \\
    NASA Perseverance      & 2021    & 8430.55 & Yg & 29500 &  88  &  0.90 \\
    \hline
    CNSA Tianwen-1         & 2021    & 8430.30 & Cd & 3000 & 250 & 1.15 \\
    \hline\hline
  \end{tabular}
\tabnote{Cd--Ceduna $30$-m (Australia); Ho--Hobart $26$-m (Australia); Hh--Hartebeesthoek $26$-m (South Africa); Mh--Mets\"{a}hovi $14$-m (Finland); T6--Tianma $65$-m (China);  Ww--Warkworth $30$-m (New Zealand); Wz--Wettzell $25$-m (Germany); Yg--Yarragadee $12$-m (Australia). CNSA--China National Space Administration, JAXA--Japan Aerospace Exploration Agency, MRO--Mars Reconnaissance Orbiter\\
$^*$The spacecraft operated in open-loop mode using the on-board oscillator during those observations.}
\end{table*}

Venus Express spacecraft was used in three different science cases by \citet{Duev2012},~\citet{Molera2014} and~\citet{Bocanegra2019}. VEX had a $24$-hour near-polar elliptical orbit with a Doppler shift ranging between $400$ and $2500\,$Hz per $19$-min scans ($0.35$ to $2.3\,$Hz/s). The spacecraft was equipped with a high-gain antenna at X-band for the radio communications. The standard SNR received with a $\sim 20$-m dish radio telescope was $10000$ ($40\,$dB). \citet{Duev2012} demonstrated precise estimates of the orbital state vectors of VEX with EVN antennas. We used analytical measurements of the topocentric frequency detections with \sw{SDtracker} to verify the model. Data were acquired with multiple $16\,$MHz frequency channels and scans of $2$ minutes. For the \sw{SWspec} data processing, we used $3.2$ million FFT points and $5\,$s integration time. These settings yielded to a frequency accuracy in the order of $50\,$mHz and an SNR of $32\,$dB. For the \sw{dPLL}, we used a $6$-degree frequency polynomial fit in the finer Doppler compensation and a $9$-degree phase polynomial to extract the residual phase of the carrier line. The final output of the carrier tone was filtered down to $1\,$Hz.

\citet{Molera2014} presented the interplanetary scintillation (IPS) studies campaign based on the investigation of VEX spacecraft carrier phase fluctuations caused by the solar wind. We used consistently a dF of $\,5\,$Hz and a dT of $\,5\,$s when running \sw{SWspec} and \sw{SCtracker}. We kept the settings constant for better comparison of the results between the stations. For the \sw{dPLL}, we used $6$-order polynomial fits to finer compensate the Doppler and the trend of the phase. The phase scintillation caused by the solar wind was evaluated in a $20\,$Hz output bandwidth.

\citet{Bocanegra2019} investigated the applicability of this setup for conducting planetary atmospheric studies by means of radio occultation experiments. For the data analysis of the occultation scans, the frequency polynomial coefficients are computed up to a few seconds before the ingress starts and a few seconds after the egress has ended. In this way, the fit corresponds to the Doppler shift due to the relative motion of the transmitter with respect to the receiver, and the remaining frequency residuals will correspond mainly to propagation effects of the signal through the target planet's atmosphere and Earth's ionosphere and troposphere. The integration time used in \sw{SWspec} and \sw{SCtracker} was $1\,$s and a $9$-order polynomial. The processing with the \sw{dPLL} was done with an integration time of $0.1\,$s and an output bandwidth of $50\,$Hz.

Mars Express spacecraft observations further developed the PRIDE techniques as presented in \citet{Duev2016},~\cite{Molera2017},~\citet{Bocanegra2018} and~\citet{Maoli2021}. MEX is in a polar orbit with a period of $7.5$ hours, which yielded to the Doppler shift measurements per scan larger that for VEX, as seen in Table~\ref{tab:observations}. The antenna-type in MEX the same as in VEX, thus, the SNR received at the stations were similar to previous experiments. \citet{Duev2016} and~\citet{Bocanegra2018} estimated the state vectors of MEX during a flyby on Phobos in $2015$. We used a combination of long ($19$-min) and short scans ($2$-min) alternating with a reference source for OD. To cope with the short scans and the fast dynamics of MEX, we used a dF of $40\,$Hz and dT of $1\,$s in \sw{SWspec} and \sw{SCtracker}. The degree of the phase polynomial used in \sw{dPLL} varied depending on the scan's orbital phase, ranging between $6$, $9$ and up to $15$ close to the actual flyby.

The IPS campaign continued in $2014$ using the MEX carrier signal. The work is currently under preparation and covers 7 years of data. Typical Doppler shift ranges from $1$ to $20\,$kHz per $19$-min scans ($1$ to $17\,$Hz/s). Due to the higher radial acceleration of the observer-spacecraft system, we needed to use lower integration times. Typical dT are $5$ or $2\,$s, while dF is $5$ or $10\,$Hz. To prioritise the consistency on the results between MEX and VEX data, we used the same order of polynomials ($6$) and output bandwidth ($20$\,Hz).

Finally,the detection of a coronal mass ejection (CME) crossing the line of sight between the MEX spacecraft and Earth was presented in~\citet{Molera2017}. Due to the large frequency fluctuations caused by the CME we had to use $1\,$s integration time with a frequency resolution of $5\,$Hz. The final output of the tone was filtered down to $20\,$Hz as usual, but we used a $8-$degree phase polynomial to de-trend the phase in \sw{dPLL}.

\subsubsection{Earth-bound satellites}

\sw{SDtracker} is also capable of processing radio signals transmitted by Earth-bound satellites. In $2011$,~\citet{Tornatore2011} made one of the first attempts to link different space geodetic techniques by observing global navigational satellites (GNSS) with VLBI radio telescopes. Observations of multiple GNSS with the VLBI radio telescopes of Medicina and Onsala were conducted as test case. The approach is similar to the work done by~\citet{Duev2012} and~\citet{Duev2016}, but with satellites that move more rapidly in the sky.

In $2015$,~\citet{Duev2015} demonstrated how to improve the orbit determination of the Spektr-R satellite of the Space VLBI mission RadioAstron on a highly elliptical orbit~\citet{Kardashev_2013}. The Doppler and phase referenced VLBI measurements obtained in this experiment were used in high-precision state vector determination of space-borne radio telescope –– a crucial component of Space VLBI.

Processing data from Earth satellites is similar to the procedure for deep space missions, however, a number of considerations must be taken into account. The Earth satellites move at much higher non–sidereal velocities on the sky as compared to deep space missions, thus, ground antennas have to either follow the target with a higher non-sidereal slewing rate or to re-point more often. Many VLBI antennas cannot continuously track non-sidereal targets. Therefore, satellite and spacecraft orbits are converted into a series of specific topocentric positions (right-ascension and declination). The antennas are thus repositioned each time in order to keep the target within their primary beam. The time interval depends on the telescope's primary beam size and the distance of the target. When observing a spacecraft orbiting Mars, the re-pointing is done every $20$ minutes. However, for GNSS targets, the re-pointing is needed every $5$ to $15$ seconds because the targets move very fast through the primary telescope beams. If the repointing rate is too slow, it can happen that the spacecraft is observed through one of the antenna's side lobes, which will result in a non-flat shape of the SNR within the scan (usually a saw-teeth pattern). The start/stop of the antenna drives during the tracking also introduce additional system noise that affects the sensitivity of Doppler measurements. Finally, one has to consider the strong transmission power emitted by these Earth orbiting satellites. Additional attenuation at the antenna may be required in order to avoid saturation of the channels with the spacecraft signal and the aliasing of harmonics in the adjacent frequency channels.

\subsection{Single-dish spectroscopy for astronomical applications}
\label{singledish}

\citet{Pogrebenko2009} conducted spectrometric observations of the Saturnian system with the VLBI radio telescopes of Medicina and Mets\"{a}hovi at the frequency of the water maser emission near $22\,$GHz. The data processing was conducted with an earlier version of \sw{SWspec}. The current version of the program can be also utilised as a single-dish spectrometer directly from any VLBI recorder.

\begin{figure}[htbp]
 \includegraphics[width=1.0\columnwidth]{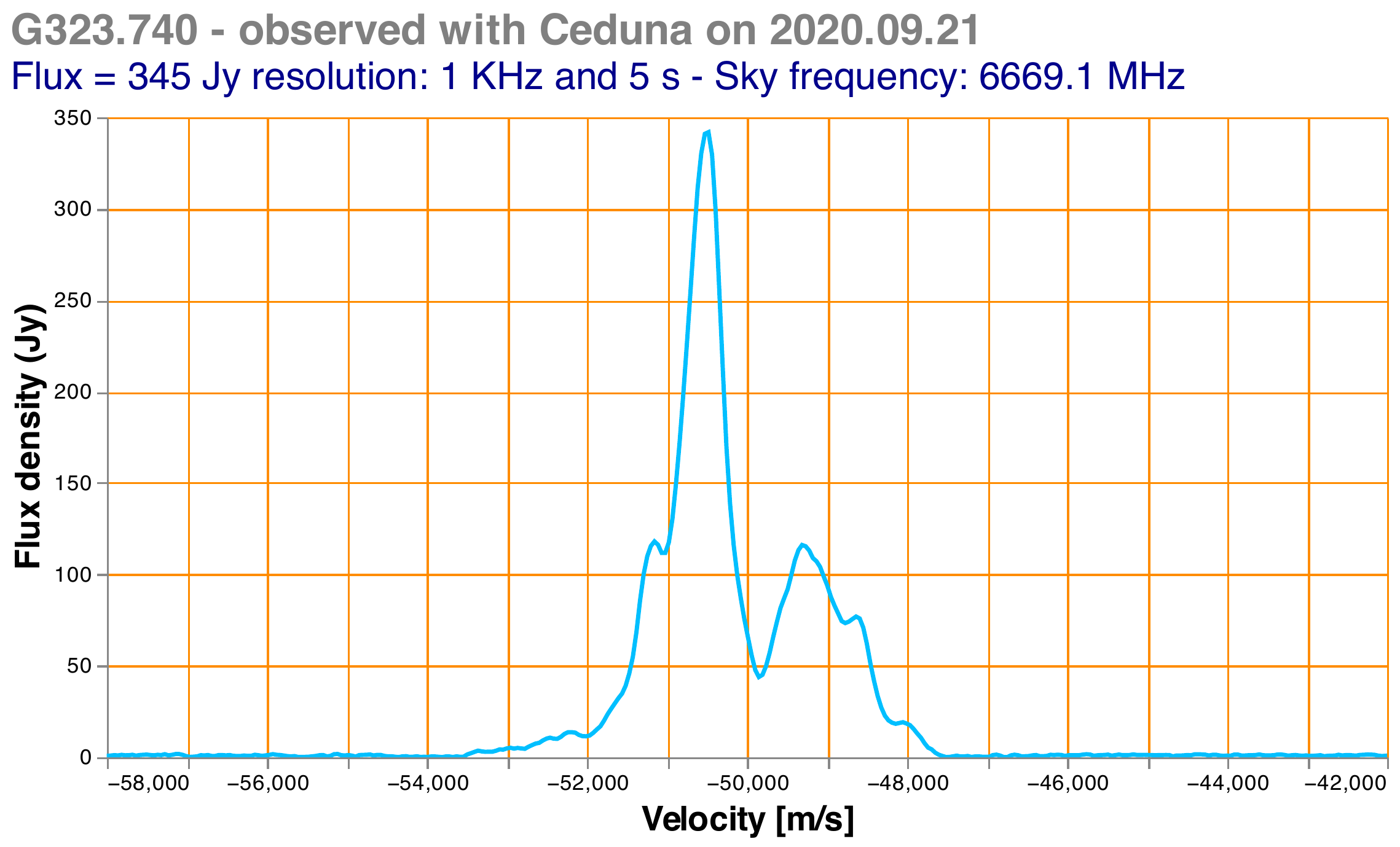}
 \caption{G$323.740$ methanol maser observed on $2020.09.21$ with the radio telescope at Ceduna (Australia) as part of the University of Tasmania's effort of long-term study of methanol masers. Spectra generated using \sw{SWspec} with $1\,$kHz frequency resolution and $5\,$s integration time.}
 \label{fig:methanol}
\end{figure}

\sw{SWspec} is used to continue the study of astronomical masers by the University of Tasmania using their network of radio telescopes. Long term monitoring of the 6.7-GHz class~II methanol maser spectral profile is necessary for the early detection of maser flares in high-mass star formation regions. These flare events allow for the detection of rare maser transitions, which can then be used to investigate the physical properties in these star-formation regions~\citet{Breen2019,Brogan2019}. In the example shown in Figure~\ref{fig:methanol}, one can see the spectral profile of the methanol maser G$323.740$ as observed with the Ceduna radio telescope on $2020.09.21$. We used \sw{SWspec} with dF of $\,1\,$kHz and dT of $\,5\,$s. The peak of the maser is at topocentric frequency of $6668.518\,$MHz. The radial velocity of the maser has been corrected for the local standard of rest.

\subsection{Bistatic radar}
VLBI-equipped radio telescopes can also be set up as a bistatic receiver in radar applications. In these operations a large antenna transmits a tone to the target of interest, while the receiver antenna(s) observes the bounced echo. The technique has been successfully applied to determine the spin states of asteroids~\citet{Busch2010} and surveying the space debris with VLBI technology~\citet{Yajima2007, Montebugnoli2010}. In $2021$, we have demonstrated the capability of the Hobart and Katherine $12$-metre antennas for tracking the echo reflected by a large space vehicle. In the experiment on $2021.03.21$ the three DSN dishes of Tidbinilla (Australia) transmitted three powerful tones to the 2015-56B rocket. These independent tones were detected at both antennas. The experiment is conducted in collaboration with NASA and it is the preparatory tests for exploring the features of VLBI radio telescopes towards Space Domain Awareness (SDA). This technique can also be applied to the planned bistatic radar observations of the ESA's JUICE mission to study Jupiter's atmosphere.

\begin{figure}[htbp]
 \includegraphics[width=1.0\columnwidth]{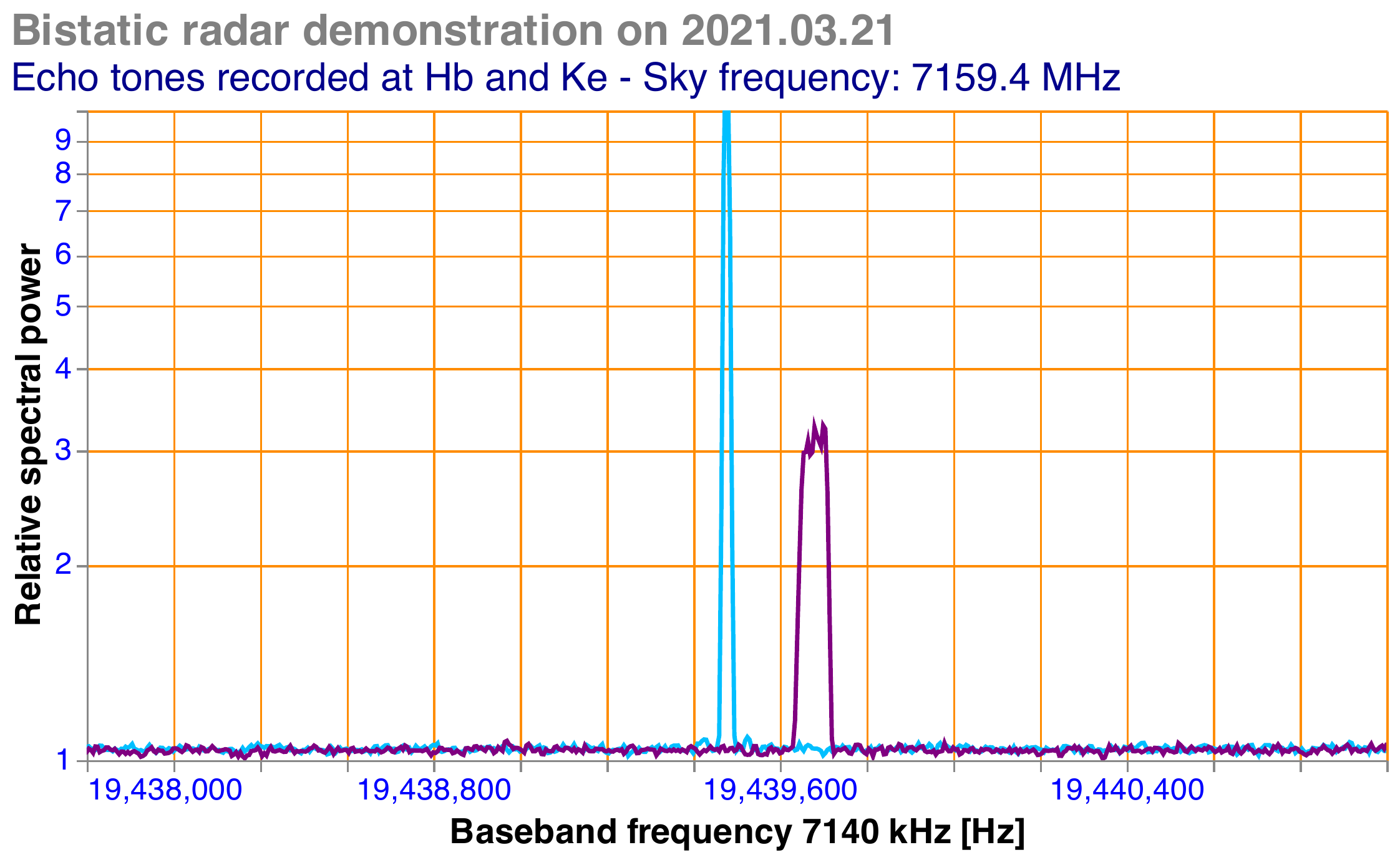}
 \caption{A single tone, transmitted by the DSN34 antenna, was reflected by the $2015-56$B space rocket and its echoes were received at Hb (blue) and Ke (purple). This bistatic radar SDA experiment was conducted on $2021.03.21$.}
 \label{fig:bistatic}
\end{figure}

Figure~\ref{fig:bistatic} shows the echoes of the signal transmitted by DSN34 as received at Ke and Hb radio telescopes. The Doppler shift in $30$ minutes was $10\,$Hz and $160\,$Hz, respectively. The Doppler shift included the movement of stations and the correction for the transmitted frequency (the so-called {\it ramp} table) to receive a static tone at Katherine.

\section{Future outlook}
\label{outlook}
A number of new experiments have been conducted recently in order to keep improving the PRIDE software pipeline. EVN observations at X-band of the NASA lander InSight on the surface of Mars have been carried out in $2020$--$21$\footnote{~\url{old.jive.nl/archive-info?experiment=ED045A\_200222}; accessed on 2021.04.20} with the aim of developing the PRIDE measurement methodology in preparation for the LaRa experiment of the ExoMars--$2022$ mission~\citet{Dehant2020}. The high precision and the multi-station Doppler data provided by PRIDE measurements can contribute into reduction of the estimated uncertainty of Mars orientation parameters. These are strongly linked to the Martian interior –– the main task of the LaRa experiment. The software presented in this paper can provide estimates of a Martian lander Doppler shift at the precision of several tens of $\mu$m/s, which are suitable for determining small variations in Mars' rotation.

In recent years, the Martian system has become a popular target for several space agencies, where multiple spacecraft and landers coexist. Simultaneous multi-spacecraft observations have many scientific applications from more accurate ephemerides’ determination to the study of gravity fields. This approach can be coupled with in-beam calibration, in which the calibrator of the VLBI observations is also observed at the same time (i.e. in the same primary beam) of the target spacecraft. This allows continuous sampling and better calibration. Another novel approach to improve calibration is the MultiView phase referenced technique that combines observations of $3$ or $4$ near-by calibrators and the target~\citet{Rioja2020}. The advantage of this technique is that reference sources can be up to $8$ degrees apart from the spacecraft. The technique can be beneficial in critical operations where strong reference sources are not near-by. Multi-spacecraft and MultiView experiments applied to PRIDE have been attempted in $2020$ and the analysis of their outcome is in progress.

An important milestone for future applications aims at exploiting observing setups in which experiments show low SNR and/or strong perturbations in the received signal. VLBI is unique in providing a combination of high sensitivity telescopes and wide geographical coverage with different local ionospheric conditions. Furthermore, the large variety of VLBI radio telescopes, which differ by their sizes and other parameters, makes possible the optimisation of PRIDE observations of a particular spacecraft in order to achieve detections with suitable SNR values. Work on the \sw{SDtracker} software is ongoing and it aims at exploiting this advantage of the PRIDE technique.

Another important software improvement currently in development focuses on spacecraft radio signals communicating at a higher frequency band. Ka-band ($32\,$GHz) is the data communication frequency adopted for the new generation deep space missions, with S-band systems becoming obsolete. ESA's BepiColombo and JUICE missions have been designed to operate at both Ka- and X-band with the high- and medium-gain antennas. The PRIDE technique is fully compatible with this new technology due to the nature of the VLBI instrumentation and the flexibility of the software presented in this work. Observations of BepiColombo, as well as Juno (see Table~\ref{tab:observations}), have already been conducted in $2020$--$21$ to test the capability of PRIDE to support JUICE mission.

\section{Conclusions}
\label{conclusions}

The Spacecraft Doppler tracking software (\sw{SDtracker}) has been developed, tested and deployed as the operational program for processing signals transmitted by spacecraft and received by Earth-based VLBI radio telescopes. It has a large number of functionalities similar to the current implementations of VLBI software correlators: auto-correlation, cross-correlation, multiple windowing, and phase calibration extraction~\citet{Deller2007}. In addition, the software package is compact, powerful and versatile. It can be installed on any standard multi-core operating system ($32$ or $64$-bit architectures). The software can process data acquired with most of the VLBI-equipped stations and VLBI data formats. The software has been updated to the latest releases of \verb!C++! compilers, Intel IPP libraries and \verb!Python3!. These upgrades and more powerful processors have reduced the data processing time by $50\,\%$ compared to the initial tests in $2010$. All data processing can be done at the station, consequently, data transfers from stations to the processing centre are no longer necessary.

\sw{SDtracker} enables estimates of the topocentric radial velocity of spacecraft with the precision of the order ten $\mu$m/s. Precise measurements of the residual phase of the carrier signal are obtained per station. This precision is relevant for supporting the goals of major science objectives of prospective planetary missions. The software has been validated in multiple experiments conducted during the past years. From precise determination of spacecraft state vectors, radio occultations experiments, gravitational measurements with flybys, to forecasting space weather. At present, the technique is being upgraded to meet requirements of the upcoming experiments with the ESA's JUICE and ExoMars--$2022$ missions.

ESA's JUICE mission adopted PRIDE as one of its scientific experiments. PRIDE will contribute into the improvement of the ephemerides of the Jovian system in support of various science tasks of the JUICE mission~\citet{Dirkx2017}. Over the $10+$ years of the mission duration, PRIDE will conduct Earth-based observations of the JUICE spacecraft to strengthen the mission science suite as already demonstrated with the tracking of other missions. Both the technique and the software developed for PRIDE are in constant evolution to include newer VLBI instrumentation and to be ready for new challenges of enhancing the scientific return of the next generation of space missions in the Solar System.

\begin{acknowledgements}
The authors would like to thank the personnel of the VLBI observatories and operational personnel of the planetary and space science missions by ESA, NASA, JAXA and Roscosmos for their support to the developments presented in this paper. Part of the experiments described in the paper have been conducted as the global VLBI project GR035 and EVN projects EM081, EC064, ED045 projects. The European VLBI Network is a joint facility of independent European, African, Asian, and North American radio astronomy institutes. A number of presented here results were obtained in observations conducted at the telescopes operated by the University of Tasmania (Australia), Mets\"{a}hovi Radio Observatory, Aalto University (Finland), INAF Institute of Radio Astronomy (Italy), Auckland University of Technology (New Zealand), Geodetic Observatory Wettzell (Germany), and the QUASAR Network, Institute of Applied Astronomy (Russia). The authors would also like to thank Paul Boven, Shinji Horiuchi, Aard Keimpema, Mark Kettenis, Mikhail Lisakov, Jon Quick, Harro Verkouter, and UTAS colleagues for their advice and support.
\end{acknowledgements}

\bibliographystyle{pasa-mnras}
\bibliography{references}
\newpage

\section*{A. APPENDIX}
\captionsetup[table]{labelformat=simple}
\setcounter{table}{0}
\renewcommand{\thetable}{A\arabic{table}}
\begin{table}[h!]
  \caption{List of control file settings available in \sw{SCtracker}.}
  \label{tab:appendixSCtrackerConfig}
  \small
  \begin{tabular}{l|l}
    \hline\hline
    \textbf{Parameter} & \textbf{Description} \\
    \hline
    InputSource & Location of the antenna raw data file \\
    ToneOffsetsFile & Separation of the DOR tones from the carrier \\
    SourceFormat & Raw file data format \\
    SinkFormat & Output data format \\
    DoSpacecraftTracking & Perform the phase correction on the data \\
    WriteDoublePrecision & Output the data in float 64 bits \\
    BitsPerSample & Input data precision \\
    SourceChannel & Number of frequency channels per frequency \\
    UseChannel & Select frequency channel to extract \\
    BandwidthHz & Input data bandwidth of the frequency channels \\
    OverSamplingFactor &  Spectral oversampling on the output spectra \\
    SourceSkipSeconds & Skip a specific number of seconds in the data \\
    PhasePolySign & Sign of the phase correction \\
    PhasePolyCoeffType & Type of the input polynomial coefficients \\
    PhasePolyOrder & Order of the input polynomial coefficients \\
    PhasePolyCpmFile & Location of the phase polynomial used as input \\
    PhasePolyCppFile & Location of the phase polynomial used as input \\
    PhaseLockSpectra & Provide statistics of the partial PLL \\
    FFTpoints & Number of spectral points to be output \\
    FFTIntegrationTime & Integration time for the spectral output \\
    FFToverlapFactor & Overlap for the spectral output \\
    PaddingFactor & Additional padding added to the spectral data \\
    WIndowType &  Window used for the spectral output \\
    WindowOutType & Window used for the output signal \\
    FilterBandwidthHz & Bandwidth of the output signal \\
    BaseFilename & Format of all the output files \\
    \hline\hline
  \end{tabular}
\end{table}

\end{document}